\documentclass[11pt,a4paper]{article}

\usepackage{physics}
\usepackage[utf8]{inputenc}
\usepackage[T1]{fontenc}    
\usepackage{slashed}
\usepackage{mathtools}        
\usepackage{bbm}              
\usepackage{bm}               
\usepackage{caption}
\usepackage{subcaption}
\usepackage{multirow}
\usepackage{array}
\usepackage{booktabs} 
\usepackage[most]{tcolorbox}

\usetikzlibrary{positioning,arrows.meta,decorations.markings,decorations.pathmorphing}

\usepackage{jheppub}          
\usepackage{graphicx}
\usepackage{comment}
\usepackage{latexsym}
\usepackage{xspace}
\usepackage{lscape}
\usepackage{color}
\usepackage{relsize}
\usepackage{tabularx}
\usepackage{amsmath}
\usepackage{amssymb}
\usepackage{lipsum}
\usepackage{feynmp-auto}
\usepackage{upgreek}
\usepackage{nicefrac}
\usepackage[titletoc]{appendix}

\usepackage{cleveref}         
\crefname{equation}{}{}
\crefname{figure}{}{}         
\crefname{table}{}{}
\crefname{section}{}{}        
\crefname{appendix}{}{}
\crefname{footnote}{}{}

\crefalias{subequation}{equation}

\def\nc{\newcommand}

\newcommand{\be}{\begin{equation}}
\newcommand{\ee}{\end{equation}}
\newcommand{\bea}{\begin{eqnarray}}
\newcommand{\eea}{\end{eqnarray}}
\newcommand{\ba}{\begin{array}}
\newcommand{\ea}{\end{array}}
\renewcommand{\vec}[1]{\bm{#1}}

\nc{\nn}{\nonumber}

\nc{\deldag}{{\mathbin{\partial\mkern-10mu/}}}
\nc{\kdag}{{\mathbin{k\mkern-10mu /}}}
\nc{\udag}{{\mathbin{u\mkern-10mu /}}}
\nc{\Ddag}{{\mathbin{D\mkern-10mu /}}}

\def\Slashnew#1{#1\kern-0.55em\raise.05ex\hbox{/}}
\def\slashnew#1{#1\kern-0.5em\raise.05ex\hbox{{$\scriptstyle /$}}}

\def\lsim{\mathrel{\raise.3ex\hbox{$<$\kern-.75em\lower1ex\hbox{$\sim$}}}}
\def\gsim{\mathrel{\raise.3ex\hbox{$>$\kern-.75em\lower1ex\hbox{$\sim$}}}}

\nc{\shalf}{\ensuremath{\textstyle \frac{1}{2}}}
\nc{\ihalf}{\ensuremath{\textstyle \frac{i}{2}}}

\def\sfrac#1#2{\ensuremath{\textstyle \frac{#1}{#2}}}

\def\emph#1{{\em #1}}

\def\lhs{{\em l.h.s.}}
\def\rhs{{\em r.h.s.}}
\def\eg{{\em e.g.}}
\def\ie{{\em i.e.}}

\def\H{{\rm H}}

\def\th{{\bar h}}

\nc{\sss}{\scriptscriptstyle}
\nc{\W}{{\sss W}}
\nc{\sparallel}{{\sss\parallel}}
\nc{\tGamma}{{\tilde \Gamma}}

\newcommand{\evec}[1]{{\bm{#1}}}              

\def\H{{\rm H}}

\def\l{{\scriptscriptstyle <}}
\def\g{{\scriptscriptstyle >}}

\nc{\CPslash}{{{\scriptscriptstyle {\rm CP}}\mkern-18mu / \mkern 10mu}}

\def\calC{{\scriptscriptstyle {\cal C}}}
\def\W{{{\cal W}}}
\nc{\HF}{{\sss \rm FH}}

\title{Quantum transport theory for neutrinos with flavor and particle-antiparticle mixing}

\author[a,b]{Kimmo Kainulainen}
\author[a,b]{Harri Parkkinen}
\affiliation[a]{Department of Physics, PL 35 (YFL), 40014 University of Jyv\"askyl\"a, Finland}
\affiliation[b]{Helsinki Institute of Physics, PL 64, 00014 University of Helsinki, Finland}
    \emailAdd{kimmo.kainulainen@jyu.fi}
    \emailAdd{harri.h.parkkinen@jyu.fi}

\abstract{We derive quantum kinetic equations for mixing neutrinos including consistent forward scattering terms and collision integrals for coherent neutrino states. In practice, we reduce the general Kadanoff--Baym equations in a few clearly justified steps to a generalized density matrix equation that describes both the flavour- and particle-antiparticle coherences and is valid for arbitrary neutrino masses and kinematics. We then reduce this equation to a simpler particle-antiparticle diagonal limit and eventually to the ultra-relativistic limit. Our derivation includes simple Feynman rules for computing collision integrals with the coherence information. We also expose a novel spectral shell structure underlying the mixing phenomenon and quantify how the prior information on the system impacts on the QKE's, leading to a direct effect on its evolution. Our results can be used for example to accurately model neutrino distributions in hot and dense environments and to study the production and decay of mixing heavy neutrinos in colliders.}

\keywords{Neutrino Mixing, Neutrino Interactions, Sterile or Heavy Neutrinos}

\arxivnumber{}
\preprint{}

\begin{document}
\maketitle

%
\section{Introduction}
%

One of the key long term goals in neutrino physics is to obtain practically solvable quantum kinetic equations (QKEs) that can accurately and consistently model the coherent neutrino flavor evolution including also decohering collisions. Traditionally neutrino oscillations have been described either in the quantum-mechanical (QM) wave-packet approach or using some partial quantum-field theory (QFT) methods. In the QM treatment~\cite{Nussinov:1976uw, Kayser:1981ye, Giunti:1991ki, Kiers:1996sh, Dolgov:1997mo, Giunti:1998ki, Cardall:2000ch, Dolgov:2002wy, Giunti:2002xg, Akhmedov:2009rb, Akhmedov:2010ms} neutrinos are described by wave packets whose widths are related to the uncertainty of neutrino momentum at the production process. In the partial QFT approaches~\cite{Akhmedov:2010ms, Bilenky:1987ty, Giunti:1993se, Grimus:1996av, Grimus:1998uh, Beuthe:2002ej, Boyanovsky:2011xq, Lello:2012gi, Grimus:2019hlq, Akhmedov:2022bjs, Kovalenko:2022goz, Krueger:2023skk, Grimus:2023ktd} neutrino production, detection and propagation are considered via compound Feynman diagrams that treat neutrinos internally as QFT propagators and externally as wave packets. In vacuum, when decoherence effects related to neutrino production, detection and propagation can be neglected, both approaches lead to the standard formula for neutrino oscillations probabilities. However, in hot and dense environments, where interactions modify neutrino flavor evolution significantly, more fundamental approaches are needed.

The kinetic theory for flavour mixing neutrinos in thermal environments was developed in early nineties~\cite{Barbieri:1989ti,Kainulainen:1990ds,Barbieri:1990vx,Enqvist:1990ad,Enqvist:1990ek,Enqvist:1991qj,Sigl:1992fn}. It led to the discovery of an activation mechanism for sterile neutrinos via mixing and decohering collisions (for the non-resonant case~\cite{Barbieri:1989ti,Kainulainen:1990ds} and for the resonant case~\cite{Barbieri:1990vx,Enqvist:1990ad,Enqvist:1990ek}), which was used to derive accurate nucleosynthesis bounds on neutrino mixing parameters~\cite{Enqvist:1991qj}. For later numerical work see also~\cite{Hannestad:2012ky,Hannestad:2015tea}. The activation mechanism of~\cite{Barbieri:1989ti,Kainulainen:1990ds,Barbieri:1990vx,Enqvist:1990ad,Enqvist:1990ek} was later used to establish popular freeze-in warm dark-matter scenarios~\cite{Dodelson:1993je,Shi:1998km}. The early derivations of the kinetic (density matrix) equations were based on the $S$-matrix formalism~\cite{Enqvist:1991qj,McKellar:1992ja} or on the operator formalism~\cite{Sigl:1992fn}. 
The resulting equations included forward-scattering (mean field) corrections and displayed the strong coupling between the particle and antiparticle sectors~\cite{Enqvist:1990ad,Enqvist:1990ek} induced by neutrino-neutrino forward scattering, which can cause lepton asymmetry instabilities in the early Universe~\cite{Enqvist:1999zs,Kainulainen:2001cb,Hannestad:2015tea,Hannestad:2013pha}. For more recent work see~\cite{Abazajian:2012ys}. Early treatments did not include the direct particle-antiparticle coherences however. These were incorporated in the CTP-based derivation in~\cite{Her082,Her081,Her083,Her09,Her10,Her11,Fid11,Jukkala:2019slc}, which resulted in general QKEs that include both the flavour and the particle-antiparticle mixing. 

Other efforts to derive quantum kinetic equations in the specific neutrino physics context include~\cite{Serreau:2014cfa,Vlasenko:2013fja,Blaschke:2016xxt,Froustey:2020mcq}, but these treatments were less general than that of~\cite{Her082,Her081,Her083,Her09,Her10,Her11,Fid11,Jukkala:2019slc}, relying on the ultra relativistic (UR) limit and the mean field limit~\cite{Serreau:2014cfa} or expansions in small perturbative quantities~\cite{Vlasenko:2013fja,Blaschke:2016xxt,Froustey:2020mcq}. These articles found some effects of the particle-antiparticle mixing dubbing them as spin coherence~\cite{Vlasenko:2013fja} or helicity coherence~\cite{Serreau:2014cfa}, and proposed that they may be relevant in hot and dense environments (see also~\cite{Volpe:2023met}). This is usually not the case however and there indeed seems to be some confusion in the literature concerning the notion of particle-antiparticle mixing. This term seems to be used when actually referring to the CP-violating {\em flavour} mixing induced effects on the evolution of the particle-antiparticle asymmetry, already included in the early treatments discussed above. Such CP-violating effects are of course at the heart of the leptogenesis~\cite{Juk21} and the electroweak baryogenesis~\cite{Kai21} problems. They are also relevant for the production and decay of heavy Majorana neutrinos in high energy physics experiments~\cite{Antusch:2020pnn} as well as for the dynamics of the supernovae explosions~\cite{Volpe:2023met}. The true particle-antiparticle mixing on the other hand, is relevant \eg~for particle production in the early universe during the (p)reheating phase after inflation~\cite{Kainulainen:2021eki,Kainulainen:2022lzp}.

Articles~\cite{Her081,Her083,Her09,Her10,Her11,Fid11} discussed dispersive corrections only at a generic level. A more detailed formulation, but still with no explicit expressions for dispersive corrections was given in~\cite{Juk21}. These corrections are important as they eventually give rise to neutrino forward-scattering potentials. A fully general and self-consistent derivation of the QKEs from fundamental principles, which include both the forward scattering potentials and the decohering collision integrals and encompass both flavour and particle-antiparticle mixing coherences, has still been missing until now. This work fills this gap responding to a demand sometimes voiced in the literature~\cite{Volpe:2023met}. Our formalism is not restricted to treating just neutrino mixing and coherences and it is indeed much more general than is necessary for most neutrino physics applications. We will still tune our presentation at all times keeping the neutrino physics motivation in mind however.

 We start from a SD-equation in the Closed Time Path (CTP) formulation~\cite{Sch61,Kel64,Cal88}. In a series of well motivated steps reduce the SD-equations to quantum kinetic equations which contain all coherence information essential for neutrino mixing and oscillations. Our derivation assumes only slowly (adiabatically) varying background fields, the validity of weak coupling expansion and eventually the spectral limit (although this is optional). We employ the local approximation that was recently developed and applied in the context of resonant Leptogenesis~\cite{Juk21} and is closely related to the coherent quasiparticle approximation (cQPA) developed in~\cite{Her081,Her083,Her09,Her10,Her11,Fid11,Jukkala:2019slc}. An essential element in our derivation is the introduction of a {\em projective representation} which reduces the original SD-equation to a set of Boltzmann-type equations for distribution functions classified according to their natural oscillation frequencies. In the weak coupling limit it also reveals a novel spectral shell-structure with new "coherence shells" that are recognized to carry information about the particle-antiparticle and flavour coherences. We generalize the work of refs.~\cite{Her082,Her081,Her083,Her09,Her10,Her11,Fid11} in several ways. In addition to using the projective representation of~\cite{Jukkala:2019slc,Juk21} we allow for an adiabatic evolution in the spatial coordinates. We also include a derivation and evaluation of the neutrino forward scattering potentials, whose general structure is much more complex than that found in~\cite{Serreau:2014cfa,Vlasenko:2013fja}. 

Our master equations take a very elegant and intuitive form of a set of Boltzmann-type equations (or a generalized density matrix equation) that fully contain the neutrino-antineutrino mixing and are straightforward to solve numerically. Our results are also valid for arbitrary neutrino masses and kinematics, while \eg~\cite{Vlasenko:2013fja,Serreau:2014cfa,Blaschke:2016xxt,Froustey:2020mcq} assume UR-limit from the outset. However, we do present our results also in the limit when particle-antiparticle mixing can be neglected and eventually in the UR-limit. Moreover, we take into account some higher (infinite) order gradient corrections in the Kadanoff--Baym-equations, which seem to be neglected elsewhere~\cite{Vlasenko:2013fja, Blaschke:2016xxt}, but are necessary for correct evaluation of the self-energies and collision terms. In addition (going also beyond the treatments in~\cite{Her082,Her081,Her083,Her09,Her10,Her11,Fid11,Jukkala:2019slc,Juk21}) we present our results clearly indicating at each step how they can be generalized to include finite width corrections beyond the spectral limit. Lastly, our derivation comes with a comprehensive set of generalized Feynman rules which provide a straightforward and systematic way to compute collision integrals for our QKEs including all coherence effects.

Our most general equations are useful to model the production of flavour-mixing particles by background fields, for example at the reheating phase after the inflation or during some phase transitions. In most other cases the particle-antiparticle mixing can be neglected however, and the frequency-diagonal limit can be assumed. This is sufficient for the resonant leptogenesis problem as well as the heavy (Majorana) neutrino-antineutrino oscillations mentioned above. Finally, the very simple UR-limit equations are useful tool to set up numerical framework to study neutrino distributions in hot and dense astrophysical environments, such as neutron stars and compact object mergers. 

The paper is organized as follows. In sections~\cref{sec:CTP,sec:decoupling} we set up the Kadanoff--Baym equations, and show how the pole- and statistical equations can be decoupled. In sections~\cref{sec:local_limit,sec:final_tas,sec:limits} we reduce these equations to local QKEs which take the form of a density matrix equation with and without the particle-antiparticle oscillations. In section~\cref{sec:collision_integrals} we show how to compute collision integrals appearing in these QKEs, and derive simple Feynman rules to automatize this task. In section~\cref{sec:General_self-energy} we compute the general 1-loop forward scattering potentials. In section~\cref{sec:weight_functions} we discuss how the localization is related to the (lack of) prior information on the system, and how in general a preparation of the system affects its evolution. Finally, section~\cref{sec:discussion} contains our conclusions and outlook.

%
\section{Kadanoff--Baym equations}
\label{sec:CTP}
%

In this section we briefly review the derivation of the Kadanoff--Baym equations for the neutrino two-point function from the CTP-formalism. Indeed, the quantity that holds the information about coherence for mixing neutrinos in out-of-equilibrium conditions is the 2-point correlation function:
\begin{equation}
\label{eq:2-point func}
iS_{{\cal C}ij}(u,v) 
\equiv \Tr{\hat{\rho} \mathcal{T}_{\cal C} [\psi_{i} (u) \bar{\psi}_{j} (v) ]},
\end{equation}
where $\hat{\rho}$ is an unknown density operator for the system, $\psi$ is the fermion field, $\mathcal{T}_\calC$ is time ordering operator and $u_0$ and $v_0$ are complex time arguments on the usual Keldysh-contour $\mathcal{C}$~\cite{Kel64}. The path ordered 2-point function $S_\calC(u,v)$ obeys the Schwinger--Dyson equation~\cite{Cho85,Lut60, Cor74, Cal88}  (we suppress the Dirac and flavor indices when there is no risk of confusion):
\begin{equation}
(S_0^{-1} * S)_{\cal C} (u,v) = \delta_{\cal C}^{4}(u - v) + (\Sigma * S)_{\cal C}(u,v),
\label{eq:Schwinger_C}
\end{equation}
where $S_0^{-1}$ is the free inverse fermion propagator, $(A*B)_{\calC}(u,v) \equiv \int_{\calC} {\rm d}^4w A(u,w)B(w,v)$ and the contour time delta function $\smash{\delta_C^{(4)}(u-v) \equiv \delta_C(u_0-v_0) \delta^3(\vec{u}- \vec{v})}$. The self-energy function $\Sigma_C$ depends on the model. It can be computed for example from the 2-PI effective action:
\begin{equation}
\label{eq:sigma var}
\Sigma_{\cal C}\equiv -i \frac{\delta \Gamma_2[S]}{\delta S(v,u)},
\end{equation}
where $\Gamma_2$ is the sum of the 2-PI vacuum graphs of the theory, truncated to a desired order in coupling constants.

The complex-time SD-equation~\cref{eq:Schwinger_C} is equivalent to a coupled set of Kadanoff--Baym equations~\cite{Kadanoff:1962book} for the real-time valued correlation functions:
\begin{equation}
\begin{split}
\big ( [S_0^{-1} - \Sigma^p] \ast S^p \big ) (u,v)  & = \delta^{(4)}(u-v) \\
\big ( [S_0^{-1} - \Sigma^r] \ast S^s  \big ) (u,v) & = \big (\Sigma^s \ast S^a \big ) (u,v).
\label{eq:KB_direct}
\end{split}
\end{equation}
Here $p=r,a$ refers to the retarded and advanced (pole) functions and $s= <,>$ to the statistical Wightman functions. For more details on precise definition of the real time functions and self-energies see~\cite{Her081,Her083,Jukkala:2019slc,Juk21}. The convolution is now defined over real time variables, and one can separate the internal and external degrees of freedom in~\cref{eq:KB_direct} by performing the Wigner transform:
\begin{equation}
g(k,x) \equiv \int \, {\rm d}^4 r \, e^{i k\cdot r} g(x + \sfrac{1}{2}r, x - \sfrac{1}{2}r),
\label{eq:Wigner}
\end{equation}
where $x \equiv (u + v)/2$ is the average coordinate and $k$ is the conjugate momentum to the relative coordinate $r=u-v$. This leads to the following mixed space equations:
\begin{equation}
\begin{split}
\hat{\slashed{K}}S^p(k,x) - \big(\Sigma^p \otimes S^p \big)(k,x) &= 1 \\
\hat{\slashed{K}}S^s(k,x) - \big(\Sigma^r \otimes S^s \big)(k,x) &=\big(\Sigma^s \otimes S^a \big)(k,x),
\label{eq:KB_mixed}
\end{split}
\end{equation}
where $\hat K \equiv k + \sfrac{i}{2}\partial_x$ and we defined a shorthand notation for the mixed space correlation function~\cite{Jukkala:2019slc}:
\begin{equation}
(\Sigma \otimes S)(k,x) \equiv e^{-\frac{i}{2}{\partial}_x^\Sigma\cdot \partial_k } 
               [ \Sigma_\mathrm{out}(\hat K, \, x) S(k,x) ].
\label{eq:wigconv}
\end{equation}
 Here the superscript $\Sigma$ indicates that the gradient $\partial^\Sigma_x$ acts only on the self-energy function, while $\partial_k$ is a total derivative. We also defined $\Sigma_{\rm out}(k,x) \equiv e^{\frac{i}{2}\partial_x^{\Sigma} \cdot \partial_k^{\Sigma}} \Sigma(k,x)$, and absorbed the mass term into the singular part of the Hermitian self-energy: $\Sigma_\H (k,x) = \Sigma_{\textrm{H,sg}}(x) + \Sigma_{\textrm{H,nsg}}(k,x)$.  The two forms of the KB-equations:~\cref{eq:KB_direct} and~\cref{eq:KB_mixed} are equivalent and exact to the given approximation for the self-energy function. Each for has its unique advantages that we shall use in what follows.

%
\section{Decoupling of the pole and the statistical equations}
\label{sec:decoupling}
%

In addition to their manifest non-locality the KB-equations~\cref{eq:KB_direct} and~\cref{eq:KB_mixed} feature a direct coupling between the statistical and the pole functions. In order to reduce them to a single local quantum kinetic equation, we must both localize them and decouple the pole equations from the statistical ones. We will address the decoupling problem first. The key idea is to split the statistical function into a background part that is strongly coupled to the pole functions and to a perturbation, whose equation formally decouples. The formal decoupling then suggests a wide range of approximate solutions that make the decoupling exact, leading to a single self-consistent equation for the perturbation.

%
\subsection{The general background solution}
%

The pole functions $S^{r,a}$ can be expressed in terms of the Hermitian function 
$S_{\rm H}$ and the spectral function $\cal A$: $S^{r,a} = S_{\rm H} \mp i\cal A$, and 
similarly $\Sigma^{r,a} = \Sigma_{\rm H} \mp i\Sigma_{\!\cal A}$. We can then write the pole equations in~\cref{eq:KB_direct} symbolically as follows:
\begin{equation}
\begin{split}
 S^{-1}_{\rm H0} *  S_{\rm H} & =\; 1 - \Sigma_{\!\cal A} * {\cal A}
\\
 S^{-1}_{\rm H0} * {\cal A} \phantom{i} & =\; \Sigma_{\!\cal A} *  S_{\rm H},
\label{eq:KB-symbolical_pole}
\end{split}
\end{equation}
while the statistical equation becomes
\begin{equation}
 S^{-1}_{\rm H0} * S^\l =\; \Sigma^\l * S_{\rm H} - i \Sigma_{\!\cal A} * S^\l + i \Sigma^\l * {\cal A}.
\label{eq:KB-symbolical_stat}
\end{equation}
The inverse Hermitian operator in these equations is defined as
\begin{equation}
 S^{-1}_{{\rm H0},\evec{k}}(x,y) \equiv \big(i\partial_x - m\big)\delta^4(x-y) - \Sigma_{\rm H}(x,y).
\label{eq:SH0inv}
\end{equation}
We chose to work explicitly with $S^<$. The equation for $S^>$ is equivalent, but not needed because ${\cal A}=\frac{i}{2}(S^\g+S^\l)$. We chose to use the direct space notation here, but one may go to Wigner space by basically just replacing "$*$" by "$\otimes$" everywhere.

We observe that while the pole functions appear explicitly in the equation for $ S^\l$, the converse is not true; statistical functions affect the pole equations only indirectly through the self-energy functions. This suggests dividing the statistical functions as $S^\l \equiv  S^\l_0 + \delta  S^\l$, where the {\em background solution} $ S^\l_0$ satisfies the equation
\begin{equation}
 S^{-1}_{\rm H0} *  S^\l_0 \equiv\;  \Sigma^\l \! *  S_{\rm H}.
\label{eq:stat0}
\end{equation}
This implies that the {\em perturbation} $\delta  S^\l$ satisfies the equation
\begin{equation}
 S^{-1}_{\rm H0} * \delta  S^\l =\; {\cal C}^\l,
\label{eq:statdeltaS}
\end{equation}
where the collision integral is given by
\begin{equation}
{\cal C}^\l=\; i\Sigma^\l * {\cal A} -i\Sigma_{\!\cal A} *  S^\l 
\;=\; \frac{1}{2}(\Sigma^\g \!*  S^\l - \Sigma^\l \!*  S^\g ).
\label{eq:collision_integral_eka}
\end{equation}

The essential feature of this construction is removing the direct coupling term $ \Sigma^\l *  S_{\rm H}$ from equation~\cref{eq:statdeltaS} for the perturbation $\delta S^\l$. We stress that $S_0^\l$ is not guaranteed to be the background solution in the sense that it would make the collision term vanish in the equation for $\delta S^\l$. For that to be true additional constraint needs to be imposed which is easiest to see by going to the Wigner space. From~\cref{eq:KB-symbolical_pole} and~\cref{eq:stat0} one readily finds the solutions  
${\cal A} =  S_{\rm H0} \otimes \Sigma_{\!\cal A} \otimes S_{\rm H}$ and 
$S^\l_0   =  S_{\rm H0} \otimes \Sigma^\l \!      \otimes S_{\rm H}$. Using these results we can write the Wigner space expression for the collision integral~\cref{eq:collision_integral_eka} as follows:
\begin{equation}
{\cal C}^\l_{\rm ad}(k,x) =  
   i\Sigma^\l \otimes S_{\rm H0} \otimes \Sigma_{\!{\cal A}} \otimes S_{\rm H}
  -i\Sigma_{\!{\cal A}} \otimes S_{\rm H0} \otimes \Sigma^\l \otimes S_{\rm H}.
\label{eq:colterm1}
\end{equation}
This expression vanishes identically if
\begin{equation}
 \Sigma^\l(k) = g^\l(k)\Sigma_{\!{\cal A}}(k)
\qquad \Rightarrow \qquad    S^\l_0(k) =  g^\l(k) {\cal A}(k),
\label{eq:consistency_condition_sigma}
\end{equation}
where $g^\l(k)$ is an arbitrary 4-momentum dependent scalar function.
Equation~\cref{eq:consistency_condition_sigma} defines a large class of consistent background solutions, including the vacuum: $g_{\rm vac}^\l(k_0) = \theta(-k_0)$ and the thermal background:%
%
%
\footnote{In the spectral and thermal limits, where ${\cal A} = \pi\epsilon(k_0)(\kdag - m)\delta(k^2-m^2)$ and $\Sigma^\l = 2f_{\rm FD}(k_0)\Sigma_{\!\cal A}$, this implies that $S^\l_0 = 2\pi\epsilon(k_0)f_{\rm FD}(k_0)(\kdag - m)\delta(k^2-m^2)$ which is the usual thermal propagator.}  
%
%
%
$g^\l_{\rm th}(k_0) = 2f_{\rm FD}(k_0/T)$. We stress that while we call $\delta S^\l$ a perturbation, it does not need to be small; we have only made a convenient division of the solutions, but no approximations yet. All equations written so far are as general as the full interacting field theory itself.

%
\subsection{Adiabatic background solutions}
%

The separation of equations~\cref{eq:KB-symbolical_pole} and~\cref{eq:stat0} from~\cref{eq:statdeltaS} suggests a way to construct efficient approximations. First note that the pole functions $S_{\rm H}$ and ${\cal A}$ are strongly constrained by the spectral sum rule, which prevents them from having rapidly varying coherence solutions~\cite{Her082,Juk21}. In many cases they can be solved in an adiabatic approximation which then must hold by construction also for the background solution $S^\l_0$. 
Taking a cue from the exact solutions for ${\cal A}$ and $S_0^\l$ used in~\cref{eq:colterm1}, we may define adiabatic solutions in the Wigner space as follows:
\begin{equation}
\begin{split}
{\cal A}_{\rm ad}
& \equiv \;  S_{\rm H0,ad}\,\Sigma_{\!{\cal A},{\rm ad}}\, S_{\rm H,ad}
\\
S^\l_{0,\rm ad}
&\equiv \,  S_{\rm H0,ad}\; \Sigma^\l_{\rm ad} \, S_{\rm H,ad},
\label{eq:KB-adiabatic}
\end{split}
\end{equation}
and $ S_{\rm H,ad} \equiv\;  S_{\rm H0,ad} (1 - \Sigma_{\!{\cal A},{\rm ad}}{\cal A}_{\rm ad})$ with $ S_{\rm H0,ad}^{-1}(k,x) \equiv \kdag - m - \Sigma_{\rm H,ad}(k,x)$\footnote{The solutions for the pole functions are equivalent with $S^p_{\rm ad}(k,x)=(\kdag-m-\Sigma^p_{\rm ad}(k,x))^{-1}$.}.

The idea is that after the division $S^\l = S^\l_{0,\rm ad} + \delta S^\l$, the term $\delta S^\l$ should describe a transient around the adiabatic background solution $S^\l_{0,\rm ad}$. Again, this is not guaranteed to hold automatically and for the consistency of the definition {\em two} additional conditions are needed. First, we interpret $\Sigma^\l \otimes S_{\rm H}$ as a coherence damping term that only gives the width to the background solution. That is, we set $\Sigma^\l \otimes S_{\rm H} \rightarrow \Sigma^\l_{\rm ad} S_{\rm H,ad}$ in equation~\cref{eq:KB-symbolical_stat} as a part of the adiabatic approximation.  Second, we  require that the collision integral~\cref{eq:collision_integral_eka} vanishes for the adiabatic solution. Dropping all gradients in Wigner space convolutions~\cref{eq:wigconv}  (note that when acting on adiabatic solution also $\hat K_0 \rightarrow k_0$), we get from~\cref{eq:colterm1}:
\begin{equation}
{\cal C}^\l_{\rm ad}(k,x) \approx 
    i\Sigma^\l_{\rm ad}  S_{\rm H0,ad} \Sigma_{\!{\cal A},{\rm ad}}  S_{\rm H,ad}
   -i\Sigma_{\!{\cal A},{\rm ad}}  S_{\rm H0,ad} \Sigma^\l_{\rm ad}  S_{\rm H,ad}.
\label{eq:colterm2}
\end{equation}
This expression vanishes if
\begin{equation}
 \Sigma^\l_{\rm ad}(k,x) = g_{\rm ad}^\l(k,x)\Sigma_{\!{\cal A},{\rm ad}}(k,x)
\quad \Rightarrow \quad    
 S^\l_{\rm ad,0}(k,x)    =  g_{\rm ad}^\l(k,x) {\cal A}_{\rm ad}(k,x),
\label{eq:consistency_condition_sigma_ads}
\end{equation}
which is the adiabatic generalization of~\cref{eq:consistency_condition_sigma}. 

Beyond the choice of $g^\l_{\rm ad}(x,k)$, there is a lot of freedom in defining the adiabatic pole solutions. It is often sufficient to work in the spectral limit, where $\Sigma_{{\cal A},\rm ad} \equiv 0$, or even in the vacuum, where $\Sigma^p_{\rm ad} \equiv 0$, but taking the finite width, dispersion and even backreaction from $\delta S^\l$ could be accounted for, as long as the constraint~\cref{eq:consistency_condition_sigma_ads} holds. There is yet more freedom in choosing the approximation for the operator $S^{-1}_{{\rm H0}}$ which can differ in the perturbation equation~\cref{eq:statdeltaS} from $S^{-1}_{{\rm H0,ad}}$ in the pole and background equations~\cref{eq:KB-symbolical_pole,eq:stat0}. Moreover, the approximations for the convolutions involving the Hermitian self-energy functions can be different from the leading order result employed for the absorptive self-energy functions leading to~\cref{eq:consistency_condition_sigma_ads}. There are thus many ways to define {\em consistent} approximation schemes for the split equations, whose {\em accuracy} will obviously depend on the problem at hand.

%
\subsection{Decoupled equation for the statistical function}
%

After an approximation scheme is defined for the pole and the background functions, the equation~\cref{eq:statdeltaS} for the perturbation $\delta S^\l$ acquires additional source terms. Given a specific scheme that satisfies~\cref{eq:KB-adiabatic} and~\cref{eq:consistency_condition_sigma_ads} and another specific definition for $S^{-1}_{\rm H0}$ to be used in~\cref{eq:statdeltaS}, we can write the decoupled equation for the $\delta S^\l$ as follows:
\begin{equation}
S^{-1}_{{\rm H0}} \otimes \delta  S^\l = {\cal S}^\l_{\rm ad} + {\cal C}^\l,
\label{eq:perturbation_eq1}
\end{equation}
where the collision term in~\cref{eq:perturbation_eq1} is written in terms of full $S^\l$ including the vanishing adiabatic part, and the source term ${\cal S}^\l_{\rm ad}$ can be written as:
\begin{equation}
{\cal S}^\l_{\rm ad} \equiv \sfrac{i}{2}\deldag S^\l_{0,\rm ad} + (\Sigma_{\rm H} - \Sigma_{\rm H, ad}) \otimes S_{0,\rm ad}^\l + g^\l (\Sigma_{\rm H} \otimes {\cal A}_{\rm ad}) - \Sigma_{\rm H} \otimes (g^\l {\cal A}_{\rm ad}).
\label{eq:sorsa1}
\end{equation}
To get this form we used the adiabatic pole equation for the spectral function along with the result~\cref{eq:consistency_condition_sigma_ads}. We remind that the approximation used for $\Sigma_{\rm H}$ does not need to be the same as the one used for the adiabatic self-energy function $\Sigma_{\rm H,ad}$ and moreover, the convolutions associated with the self-energies can be computed also to higher order in gradients. Indeed, note that the last two terms cancel to the lowest order in gradients.

The gradient source $\sfrac{i}{2}\deldag S^\l_{0,\rm ad}$ is relevant in applications with rapidly changing backgrounds, such as resonant leptogenesis~\cite{Juk21}. On the other hand, in the transport equations for the electroweak baryogenesis the entire CP-violating source term comes from the two last terms in equation~\cref{eq:sorsa1}~\cite{Kai21}. However, both terms can be dropped in very slowly varying backgrounds which is usually the case in light neutrino physics. The term involving $\Sigma_{\rm H}$ on the other hand, will allow accounting for the forward scattering corrections to the evolution equations even when $\Sigma_{\rm H,ad}$ was set to zero in the pole and the background equations. Conversely, if $\Sigma_{\rm H,ad} = \Sigma_{\rm H}$, then the self-energy corrections are already resummed to the quasiparticle dispersion relation and also the explicit self-energy corrections drop from the source term. We will return to this issue in section~\cref{sec:quasiparticle_basis}.

It will be useful to rewrite equation~\cref{eq:perturbation_eq1} in an alternative form, in terms of the full $S^\l$-function:
\begin{equation}
S^{-1}_{{\rm H0}} \otimes S^\l = \Sigma^\l_{\rm ad} S_{\rm H,ad} + {\cal C}^\l.
\label{eq:perturbation_eq2}
\end{equation}
The first term on the right hand side drops in the spectral limit, leaving no explicit source terms in~\cref{eq:perturbation_eq2}. The gradient source and the sources arising from the different approximations imposed on the Hermitian self-energy functions in~\cref{eq:perturbation_eq1} remain however, hidden in the notation. 

%
\section{Local equations}
\label{sec:local_limit}
%

The quintessential feature of the KB-equations~\cref{eq:KB_direct,eq:KB_mixed}, and of~\cref{eq:perturbation_eq1,eq:perturbation_eq2} is their {\em non-locality}. Our quest to reduce~\cref{eq:perturbation_eq2} to a {\em local} density matrix equation then clearly requires some further approximations. From the Wigner-space point of view the task is to curtail the infinite expansions in gradients. This can be justified by a further adiabaticity assumption, now concerning the perturbation $\delta S^\l$, or it can be enforced by integration over some of the momentum variables (encoding the lack of information on them). In what follows, we shall use both methods.

%
\subsection{Adiabaticity in space coordinates}
%

In essentially all problems relevant for neutrino physics, backgrounds are changing slowly in microscopic scales. This means that we can drop all spatial gradients acting on self-energies in the Wigner space evolution equations. In a generalization from the purely homogeneous case studied in~\cite{Juk21}, we keep the gradients acting on the correlation function however. To this end we Wigner transform equation~\cref{eq:perturbation_eq1} only over the spatial coordinates and then work in the adiabatic limit. The result is the following 2-time equation:
\begin{equation}
     \Big(i\partial_{t_1} + \sfrac{i}{2}\evec{\alpha}\cdot\nabla - {\cal H}_\evec{k}\Big)
       \bar S^\l_\evec{kx}(t_1,t_2) 
       - (\bar\Sigma_{{\rm H}\evec {kx}} * \bar S_\evec{kx}^\l)(t_1,t_2)
       = i\bar {\cal C}^\l_\evec{kx}(t_1,t_2),
\label{eq:KB-adiab2time}
\end{equation}
where $\bar {\cal C}^\l_\evec{kx} \equiv \gamma^0 {\cal C}^\l_\evec{kx}\gamma^0$. The convolution is now only over time and we identified the vacuum Hamiltonian
\begin{equation}
\label{eq:matter Ham}
\mathcal{H}_{\vec{k} ij} = \delta_{ij} (\bm{\alpha} \cdot \vec{k} \, + m_i\gamma^0),
\end{equation}
with $\bm{\alpha} \equiv  \gamma^0 \bm{\gamma}$ and the correlation functions and self-energies with over-bars are defined as $\bar S^s \equiv iS^s\gamma^0$, $\bar S^p \equiv S^p\gamma^0$ and $\bar \Sigma^s \equiv i\gamma^0\Sigma^s$, $\bar \Sigma^p \equiv \gamma^0\Sigma^p$. Also, to keep our notation compact and to  highlight the essential features related to the time variable, we defined a shorthand notation $\bar S^\l_\evec{kx}(t_1,t_2)\equiv \bar S^\l(\evec{k},\evec{x};t_1,t_2)$. Finally, we dropped the explicit source term appearing in~\cref{eq:perturbation_eq2}, because we are eventually going to the spectral limit with the adiabatic solutions. This source could be simply added at any point of the derivation however.

%
\subsection{Localization in time}
%

Our next step is to localize equation~\cref{eq:KB-adiab2time} in time. From the 2-time perspective the motivation for this is obvious as discussed above. From the Wigner space point of view the localization corresponds to an integration over the frequency variable%
%
%
\footnote{The local function $S_\evec{kx}(t,t)$ is the lowest frequency moment of the Wigner-space function $S(k,x)$:
\begin{equation}
S_\evec{kx}(t,t) = \lim_{t'\rightarrow t} S(\evec{k};\evec{x},t,t')
= \lim_{t'\rightarrow t} \int \frac{{\rm d}k_0}{2\pi} e^{-ik_0(t'-t)} S(k,x) 
= \int \frac{{\rm d}k_0}{2\pi} S(k,x).
\label{eq:localS}
\end{equation}
}.  
%
%
In other words, localization corresponds to working to the lowest order in a moment expansion in the frequency variable in Wigner space. We will discuss localization from a more general point of view in section~\cref{sec:weight_functions}, relating it to a definite statement of the prior information on the system.

Deriving local equation is simple. We start by taking the total derivative of $S^\l_\evec{kx}(t,t)$ and using the chain rule to get
\begin{equation}
    \partial_t S^\l_\evec{kx}(t,t) =
    \frac{\rm d}{{\rm d} t} \bigl[S^\l_\evec{kx}(t,t)\bigr]
    = \lim_{t_1,t_2\rightarrow t}\bigl(\partial_{t_1} S^\l_\evec{kx}(t_1,t_2) + \partial_{t_2} S^\l_\evec{kx}(t_1,t_2) \bigr) \text{.}
    \label{eq:total_derivative_of_local_S}
\end{equation}
Equation~\cref{eq:total_derivative_of_local_S} can then be combined with~\cref{eq:KB-adiab2time} and its  Hermitian conjugate to obtain the local equation:
\begin{equation}
\begin{split}
\phantom{Ha}
    \partial_t {\bar S}_\evec{kx}^\l (t,t) 
    & + \sfrac{1}{2} \big\{ \evec{\alpha}\cdot\nabla ,\, \bar S^\l_\evec{kx}(t,t) \big\}
    + i\big[ {\cal H}_\evec{k} ,\, \bar S^\l_\evec{kx}(t,t) \big]
    \\
     & + i(\bar\Sigma_{{\rm H}} * \bar S^\l)_\evec{kx}(t,t)
       - i(\bar S^\l * \bar\Sigma_{\rm H})_\evec{kx}(t,t) 
      = \big( \bar {\cal C}^\l_{\evec{kx}}(t,t) + h.c. \big).
    \label{eq:exact_equation_for_local_S}
\end{split}
\end{equation}
A time-dependent Hamiltonian can be induced by adding singular terms to the self-energy function $\Sigma_{\rm H}$. Likewise, the source term appearing in~\cref{eq:perturbation_eq2} could be easily added to complete the finite width corrections for the background solution.

Equation~\cref{eq:exact_equation_for_local_S} is local by construction but it is no longer closed. This becomes evident when one writes explicitly any of the local convolutions appearing in~\cref{eq:exact_equation_for_local_S}: 
\begin{equation}
\phantom{Ha}
(\bar \Sigma *\bar S)_\evec{kx}(t,t) 
= \int \! {\rm d} w_0 \,\bar \Sigma_\evec{kx}(t,w_0) \bar S_\evec{kx}(w_0,t)
  \approx \int \! \frac{{\rm d} k_0}{2\pi} \,\bar\Sigma(\hat K_0,\evec{k},x) \bar S(k,x).
\label{eq:local_convolution}
\end{equation}
In the second step we used~\cref{eq:wigconv}, dropped the total derivatives under the integral and used the adiabatic assumption to replace $\bar \Sigma_{\rm out} \rightarrow \bar\Sigma$. From the 2-time perspective the problem is that the convolution depends on the unknown $\delta \bar S_\evec{kx}(w_0,t)$ for arbitrary $w_0$, while equation~\cref{eq:exact_equation_for_local_S} only yields the solution for $w_0=t$. The loss of closure is a generic problem in deriving Boltzmann type equations from the SD-equations, 
and the usual solution is to reduce the Wigner space correlation functions to spectral limit~\cite{Her081,Juk21}. We will also go to spectral limit eventually, but for now we continue to develop the more general formalism including finite width corrections. To facilitate this we next introduce a novel homogeneous decomposition~\cite{Juk21} for the perturbation $\delta S^\l$.

%
\subsection{Homogeneous Ansatz}
\label{sec:homogeneous_Ansatz}
%

Remember that equation~\cref{eq:exact_equation_for_local_S} is really an equation for $\delta \bar S^\l$, even though we wrote it in terms of the full $\bar{S}^\l$ for the simplicity of notation. Equations for $\delta \bar S^\l$ admit a broader class of solutions than do the pole equations~\cref{eq:KB-symbolical_pole} and the equation~\cref{eq:stat0} for the background. While the latter two admit only {\em inhomogeneous} solutions (in the sense of classifying the solutions to differential equations), equation~\cref{eq:statdeltaS} and its descendants have also {\em homogeneous} solutions which can always be written in the following form~\cite{Juk21}:
\begin{equation}
    \delta\bar  S^\l_\evec{kx} (t;u_0, v_0)  = 2 \bar{\cal A}_\evec{kx}(u_0,t)
    \,\delta\bar S^\l_\evec{kx}(t,t)\, 2 \bar{\cal A}_\evec{kx}(t, v_0).
    \label{eq:asa1}
\end{equation}
Homogeneous solutions are often used to model transients set up at some initial time $t_{\rm in}$. Indeed, the spectral function is the unitary time evolution operator in the free theory limit: $2\bar {\cal A}_0(t_1,t_2) = U(t_1,t_2)$, whose correct normalization is ensured by the spectral sum rule $2\bar {\cal A}(t,t) = \mathbbm{1}$. Here we follow the idea of~\cite{Her10,Her11,Fid11,Juk21} and use~\cref{eq:asa1} as an {\em ansatz} for the non-local terms in the convolution integrals. This makes sense in dissipative systems with a finite damping rate $\gamma_{\evec k}$ where only points with $|u_0-v_0| \lsim \gamma_{\evec k}$ are strongly correlated; for more discussion see~\cite{Juk21}. Remarkably, the structure~\cref{eq:asa1} transforms all convolutions in~\cref{eq:statdeltaS} into simple matrix products:
\begin{equation}
(\bar \Sigma * \delta \bar S^s)_\evec{kx}(t,t) 
= (\bar \Sigma * 2\bar {\cal A})_\evec{kx}(t,t)\, \delta \bar S^s_\evec{kx}(t,t)
\equiv  \bar \Sigma_{{\rm eff},\evec{k}}(x) \, \delta \bar S^s_\evec{k}(x).
\label{eq:sigma_eff}
\end{equation}
Here we used the spectral sum rule and we also adopted the more convenient notation setting \eg~$\delta S^\l_\evec{kx}(t,t) \equiv \delta S^\l_\evec{k}(t,\evec{x}) \equiv \delta S^\l_\evec{k}(x)$. The reverse ordered convolutions are just the Hermitian conjugates of the above: $(\delta \bar S^s* \bar \Sigma)_\evec{kx}(t,t)  = [(\bar \Sigma * \delta\bar S^s)_\evec{kx}(t,t)]^\dagger$.

%
\subsection{The Master equation}
\label{sec:master_equation}
%

The result~\cref{eq:sigma_eff} allows a tremendous simplification, reducing the original integro-differential equations to a set of ordinary differential equations. Indeed we can now immediately recast our local equation~\cref{eq:exact_equation_for_local_S} simply as:
\begin{equation}
    \partial_t {\bar S}_\evec{k}^\l 
    + \sfrac{1}{2}\{\evec{\alpha}\cdot\nabla, \, {\bar S}_\evec{k}^\l\}
    = -i\big[{\cal H}_\evec{k} ,  \bar S^\l_\evec{k} \big]
     - i \Xi^\l_\evec{k} + \bar {\cal C}^\l_{{\rm H},\evec{k}},
    \label{eq:master}
\end{equation}
where the forward scattering term is defined as
\begin{equation}
\Xi^\l_\evec{k} = \bar\Sigma^{\rm H}_{{\rm eff}\evec{k}} \delta \bar S^\l_\evec{k} 
                + \bar \Sigma_{{\rm H},\evec{k}} * S_{0\evec{k}}^\l \; - \;  h.c., 
\label{eq:master_forward}
\end{equation}
and the Hermitian part of the local collision integral is 
\begin{equation}
\bar{\cal C}^\l_{{\rm H},\evec{k}} \equiv - \frac{1}{2}  
    \Big(\bar\Sigma^\g_{{\rm eff},\evec{k}} \bar S^\l_\evec{k}
       - \bar\Sigma^\l_{{\rm eff},\evec{k}} \bar S^\g_\evec{k} + h.c. \Big).
\label{eq:master_coll}
\end{equation}
We suppressed the $x$-dependence in all correlation and self-energy functions for clarity and we remind that we dropped the source term appearing in~\cref{eq:perturbation_eq2}. 

Equation~\cref{eq:master} is our final master equation. While the effective self-energies are still convolutions with the spectral function, they no longer depend explicitly on the perturbation. Also the convolution $\bar \Sigma_{{\rm H},\evec{k}} * S_{0\evec{k}}^\l$ is just some known function. In the spectral limit it can be absorbed into the $\delta \bar S^\l_\evec{k}$-term in~\cref{eq:master_forward}, which then allows writing $\Xi^\l_\evec{k} \rightarrow  \bar\Sigma^{\rm H}_{{\rm eff}\evec{k}} \bar S^\l_\evec{k} - h.c.$ The master equation is formally closed, assuming that self-energy functions are defined externally. We define the Hermitian self-energy function in sections~\cref{sec:sub_spectral_adiabatic} and~\cref{sec:General_self-energy} and in section~\cref{sec:collision_integrals} we show that~\cref{eq:asa1} allows expressing all self-energies including those in the collision integral in terms of the local correlation function. Note that no specific approximation scheme for the pole and the background solutions was needed in the above derivation. Different choices would lead to slightly different effective self-energy functions without changing the form of the master equation. 

%
\section{Density matrix equations}
\label{sec:final_tas}
%

We now proceed to derive the quantum kinetic (density matrix) equations from the master equation~\cref{eq:master}. These equations take the simplest form in a projective representation onto the helicity and the Hamiltonian eigenbases. This classifies solutions according to their eigenfrequencies which simplifies equations and helps the analysis and interpretation. We will work explicitly in the vacuum Hamiltonian basis, as this is sufficient in most, if not all problems in neutrino physics. A generalization to the quasiparticle basis incorporating a resummation of thermal corrections to the Hamiltonian is discussed in section~\cref{sec:quasiparticle_basis}.

\paragraph{Vacuum representation}

In a homogeneous and isotropic system helicity is conserved and the whole Dirac algebra is spanned by eight primitive structures that we list in~\cref{eq:basis1}. However, these structures can be more conveniently chosen~\cite{Jukkala:2019slc,Juk21} in the following combinations:
\begin{equation}
 \label{eq:total pro ope}
  P_{\vec{k} h i j}^{ab} = N_{\vec{k}ij}^{ab} P_{\vec{k} h} P_{\vec{k} i}^a \gamma^0 P_{\vec{k} j}^{b},
\end{equation}
where the helicity%
%
%
\footnote{The non-covariant operator $P_{\evec{k}h}$ corresponds to helicity only in the rest frame of the energy eigenstate. Otherwise it measures the spin along the momentum in a given frame. The covariant helicity operator $P_h = \frac{1}{2}(1+\gamma^5 \slashed{s}_i)$ with $s_i= (\abs{\vec{k}}, \omega_{\evec{k} i} \hat{\vec{k}})/m_i$ cannot be used, because its definition would require an exact definition of the energy shell, and this information is not available to us. Indeed, if it were, then each state would be defined precisely and no oscillations would take place.} and the vacuum energy projection operators are defined as: 
%
%
\begin{equation}
P_{\vec{k} h} \equiv \frac{1}{2} 
   \big( \mathbbm{1} + h \bm{\alpha} \cdot \hat{\vec{k}} \gamma^5 \big )
\qquad \textrm{and} \qquad 
P_{\vec{k} i}^{a} \equiv \frac{1}{2} 
   \big( \mathbbm{1} + a \frac{\mathcal{H}_{\vec{k} i}}{\omega_{\vec{k} i}} \big ).
\label{eq:projection ope}
\end{equation}
Here $h= \pm1$ is the helicity, $a,b=\pm1$ are the energy sign indices and $\smash{\omega_{\vec{k} i}= (\bm{k}^2+m_i^2)^{1/2}}$ is the vacuum energy of the neutrino eigenstate. Projection operators satisfy completeness relations $P_{\vec{k}i}^+ +P_{\vec{k}i}^- = P_{\vec{k}+} +P_{\vec{k}-} = \mathbbm{1}$, the orthogonality and idempotence relations $P_{\vec{k}i}^{a} P_{\vec{k}i}^{b} = \delta_{a b} P_{\vec{k}i}^a $ and $P_{\vec{k} h} P_{\vec{k} h'} = \delta_{h h'} P_{\vec{k} h}$, and the eigenequations $ \mathcal{H}_{\vec{k}i}  P_{\vec{k}i}^{a} = P_{\vec{k}i}^{a} \mathcal{H}_{\vec{k}i} = a \omega_{\vec{k} i} P_{\vec{k}i}^{a}$ as well as $\bm{\alpha} \cdot \hat{\vec{k}} \gamma^5 P_{\vec{k} h}  = P_{\vec{k} h} \bm{\alpha} \cdot \hat{\vec{k}} \gamma^5  = h P_{\vec{k}i}^{a}$. The normalization factors $N_{\vec{k}ij}^{a b}$ are most conveniently chosen as follows:
\begin{equation}
    \begin{split}
    N_{\vec{k} ij}^{ab} \equiv \Big({\rm Tr}\big[P_{\vec{k} h} P_{\vec{k} i}^{a} \gamma^0  P_{\vec{k} j}^{b} \gamma^0 \big]\Big)^{-1/2} = \Big[\frac{2 \omega_{\vec{k} i}\omega_{\vec{k} j}}{\omega_{\vec{k} i}\omega_{\vec{k} j} + ab (m_i m_j - \abs{\vec{k}}^2)} \Big]^{1/2}.
    \end{split}
    \label{eq:normalization_fac}
\end{equation}
This construction generalizes directly to the adiabatic case, so that we can parametrize our Wightman functions without any loss of generality as follows:
\begin{equation}
\bar S_{\evec{k} ij}^\l(t,\evec{x}) = \sum_{h a a'}  f_{\evec{k} h ij}^{\l aa'}(t,\evec{x}) 
P_{\vec{k} h i j}^{a a'},
\label{eq:correlator_par}
\end{equation}
\vskip-0.2 cm \noindent
where $\smash{f_{\vec{k} h ij}^{\l a a'}(t,\evec{x})}$ are some yet unspecified functions. Given the normalization~\cref{eq:normalization_fac} it is easy to show that ${\rm Tr}[\bar S^\l_{\bm{k}ij}P_{\evec{k}hji}^{e'e}] = f^{\l ee'}_{\bm{k}hij}$. This normalization also simplifies the dynamical equations and leads to the standard normalization of the distribution functions in the thermal limit.

%
\subsection{Projected master equation}
\label{sec:projected_master}
%

Using the projective representation~\cref{eq:normalization_fac} we can easily derive a generalized density matrix form for the master equation~\cref{eq:master}. We insert~\cref{eq:normalization_fac} into~\cref{eq:master}, multiply with $\smash{P_{\vec{k} h j i}^{e'e}}$ (note the order of the flavour and the energy sign indices) and take a trace over the Dirac indices to extract scalar equations for the eigenfunctions $\smash{f_{\vec{k} h ij}^{\l e e'}(t,\evec{x})}$%
%
%
\footnote{We implicitly assumed that in the forward scattering terms in~\cref{eq:AH4} the background solution has the same form as the perturbation. To be consistent with our most general assumptions, one should replace \eg
\begin{equation}
(\W^{{\rm H}ee'}_{\evec{k}hij})^{l}_{a} f_{\vec{k} h lj}^{\l a e'} \rightarrow
(\W^{{\rm H}ee'}_{\evec{k}hij})^{l}_{a} \delta f_{\vec{k} h lj}^{\l a e'} +
\Tr[P_{\vec{k} h ji}^{e'e} (\bar{\Sigma}_{\rm H} \otimes S^\l_{0})].
\nonumber
\end{equation}
While the trace-term is a known function, it has a different form than the first term, except in the spectral limit. Remember that we also dropped the source term appearing in~\cref{eq:perturbation_eq2}, which would add a further term: $\smash{{\rm Tr}[( (\Sigma^\l_{{\rm ad},\evec{k}h} S_{{\rm H, ad},\evec{k}h})_{ij} + h.c.)  P_{\vec{k} h ji}^{e' e}]}$ to the \rhs~of~\cref{eq:AH4}. Instead of writing equation~\cref{eq:AH4} in a complete but cumbersome form for $\delta f^{\l ee'}_{\bm{k}hij}$, we prefer the simpler, slightly inaccurate notation, keeping these caveats in mind. These issues are eventually not relevant for the neutrino physics applications where the spectral limit can be taken.}: 
%
%
\begin{equation}
    \boxed{
        \begin{aligned}
            \partial_t  f_{\vec{k} h ij}^{\l e e'} + ({\cal V}_{\bm{k} h ij}^{e'e})_{aa'} \hat{\bm k} \cdot \bm{\nabla}  f_{\vec{k} h ij}^{\l a a'} = & -2i\Delta\omega_{\bm{k} ij}^{e e'}  f_{\vec{k} h ij}^{\l e e'} 
            + \Tr[ \bar{\mathcal{C}}^\l_{{\rm H},\vec{k}h ij}   P_{\vec{k} h ji}^{e' e}] 
            \\
            & - i(\W^{{\rm H}ee'}_{\evec{k}hij})^{l}_{a} f_{\vec{k} h lj}^{\l a e'}
            + i[(\W^{{\rm H}e'e}_{\evec{k}hji})^{l}_{a}]^* f_{\vec{k} h il}^{\l e a},
        \label{eq:AH4}
        \end{aligned}
    }
\end{equation}
where $\hat{\bm k} = {\bm k}/|{\bm k}|$, a sum over the repeated indices $a$ and $l$ is understood and e defined the oscillation frequency:
\begin{equation}
2\Delta \omega_{\vec{k} ij} ^{ee'} \equiv \omega^e_{\vec{k}i} - \omega^{e'}_{\vec{k}j},
\label{eq:shell-frequencies}
\end{equation}
with $\omega^e_{{\bm k}i} \equiv e\omega_{{\bm k}i}$. The forward scattering coefficient tensor is given by:
\begin{equation}
(\W^{{\rm H}ee'}_{\evec{k}hij})^{l}_{a}  \equiv 
\Tr[P_{\vec{k} h ji}^{e'e} \bar{\Sigma}^{\rm H}_{{\rm eff}\vec{k}il} P_{\vec{k} hlj}^{a e'}].
\end{equation}
In deriving~\cref{eq:AH4} we used $(f_{\vec{k} h il}^{\l ae})^* = f_{\vec{k} h li}^{\l e a}$, which follows from the Hermiticity of $\bar S^\l_\evec{k}(t,t)$, as well as the orthogonality relation ${\rm Tr}[P_{\vec{k} h ij}^{a a'} P_{\vec{k} h ji}^{e'e}] = \delta_{a'e'}\delta_{ae}$ and we defined

\begin{equation}
\Tr[ P_{\vec{k} h ji}^{e' e} \evec{\alpha} P_{\vec{k} h ij}^{a a'}] 
+\Tr[ P_{\vec{k} h ij}^{a a'} \evec{\alpha} P_{\vec{k} h ji}^{e' e}] 
\equiv ({\cal V}_{\vec{k} h ji}^{e'e})_{aa'} \evec{k},
\label{eq:easy tra}
\end{equation}
where the velocity tensor then is 
\begin{equation}
({\cal V}_{\vec{k} h ij}^{e'e})_{aa'} = \delta_{a'e'} {\cal V}_{\vec{k} h ij}^{eae'}
+ \delta_{ae} {\cal V}_{\vec{k} h ji}^{a'e'e},
\label{eq:isoD-def}
\end{equation}
with
\begin{equation} 
  {\cal V}_{\vec{k} h ij}^{a b c} \equiv \frac{1}{2} N_{\vec{k} ij}^{ac} N_{\vec{k} ij}^{bc}  
  \Big( v_{{\bm k}i} \Big[ \frac{a}{(N_{\vec{k} ij}^{bc})^2} 
  + \frac{b}{(N_{\vec{k} ij}^{ac})^2}\Big] - v_{{\bm k}j}c\delta_{a,-b} \Big),
  \label{eq:Dtensor}
\end{equation}
and $v_{{\bm k}i} \equiv |{\bm k}|/\omega_{{\bm k}i}$.

The master equation~\cref{eq:AH4} is written in terms of frequency states rather than particle and antiparticle solutions. The positive frequency solutions directly correspond to particles of course, while the antiparticles correspond to the negative frequency solutions with inverted 3-momenta. At the level of distribution functions this implies the following relation:\footnote{
%
%
The minus sign in this definition is a convention that arises from our definition of the spectral representation~\cref{eq:correlator_par}. Indeed, from $\bar S^\g + \bar S^\l = 2\bar {\cal A}$ we get $f^{\g ab}_{\bm{k}hij} + f^{\l ab}_{\bm{k}hij} = a\delta_{ab}\delta_{ij}$. For positive frequencies the usual relation $f^{\g}_{\bm{k}hij} = \delta_{ij}-f^{\l}_{\bm{k}hij}$ then arises with  $f^<_{\bm{k}hij} \equiv f^{<++}_{\bm{k}hij}$. For the negative frequencies however, $f^{\g--}_{\bm{k}hij} = -\delta_{ij}-f^{\l--}_{\bm{k}hij}$ and the minus sign in~\cref{eq:FS} is needed to give the correct relation $\bar f^\g_{\evec{k}hij} = \delta_{ij} - \bar f^\l_{\evec{k}hij}$.
}
%
%
\begin{equation}
\bar f^{\l,\g}_{\evec{k}hij} = -f^{\g,\l --}_{(-\evec{k})hij},
\label{eq:FS}
\end{equation}
where functions in the \lhs~refer to antiparticles. It is indeed notationally much simple to work with frequencies without an explicit identification of antiparticles at the level of the evolution equation~\cref{eq:AH4}. One can always convert the initial conditions and the final results to the particle-antiparticle language using~\cref{eq:FS}, however.

The first term on the right hand side of~\cref{eq:AH4} comes from the commutator with ${\cal H}_{\evec k}$ and it induces the leading time-dependence of solutions according to~\cref{eq:shell-frequencies}. The left-hand side of~\cref{eq:AH4} displays a modified Liouville term where the tensor $({\cal V}_{\vec{k} h ij}^{e'e})_{aa'}$ encodes the effect of different group velocities on the coherence evolution. The interaction terms are cleanly separated into a collision integral and the forward scattering terms. All terms in~\cref{eq:AH4} thus have a clear physical meaning and the apparent complexity of the equation merely reflects its wide generality; equation~\cref{eq:AH4} describes both flavour and particle-antiparticle oscillations for arbitrary neutrino masses  with arbitrary interactions in backgrounds that are only constrained to be adiabatic in space, with a large freedom in the choice of the adiabatic approximation scheme. 

To proceed we must finally specify our approximation scheme(s) and define how to compute the self-energy functions and collision integrals. We will first consider the simplest approximation, \ie~the (vacuum) spectral limit. After that we will define forward scattering terms involving known Hermitian self-energy structures. The definition and evaluation of generic self-energies and the collision integral will be discussed in section~\cref{sec:collision_integrals}.

%
\subsection{Spectral limit}
\label{sec:spectral_limit}
%

\begin{figure}[t]
    \centering
    \includegraphics[width=0.9\textwidth]{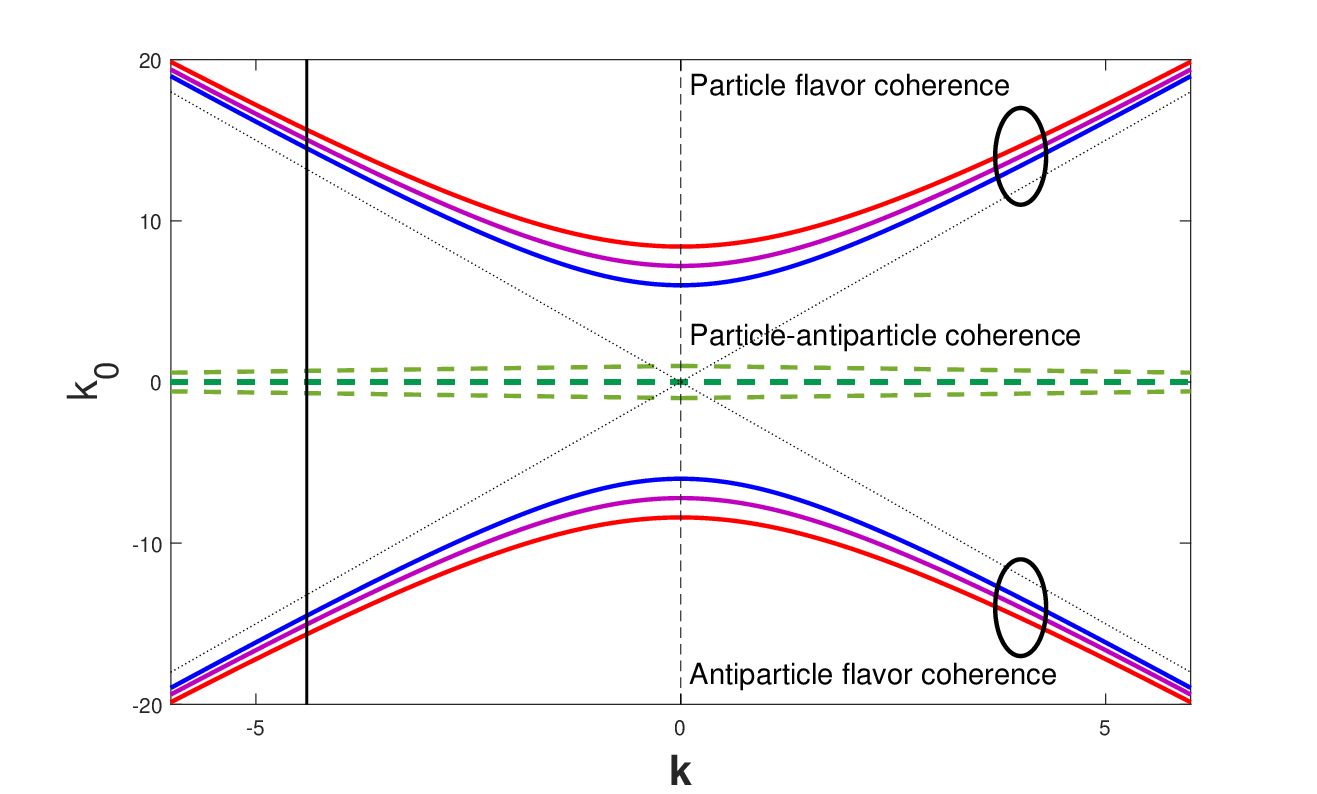}
    \caption{Shown is the cQPA-shell structure for two-neutrino mixing. (The continuation to negative ${\bm k}$ is a simplification of the full 3-dimensional rotation symmetry.) Blue and red lines denote the mass shells, purple lines show the flavor coherence shells, and green dashed lines are the particle-antiparticle coherence shells. Black ellipses illustrate the uncertainty on momentum and frequency in the preparation of the system, which would prevent determining exact flavour shells (mass-eigenstates), but would eliminate the particle-antiparticle mixing. The black line on the negative momentum side illustrates the use of a flat weight in frequency and ideal accuracy in momentum (see section~\cref{sec:weight_functions}).} 
    \label{fig:shell structures}
\end{figure}

Spectral solutions are easy to find directly using the projective parametrization~\cref{eq:correlator_par}. One can start with the Hermitian part of the statistical KB-equation~\cref{eq:KB_mixed} in the collisionless limit to the zeroth order in gradients: 
\begin{equation}
 2k_0 \bar{S}^s(k,x) = \{ \mathcal{H}_{\vec{k}},\bar{S}^s(k,x)\}. 
\end{equation}
Inserting here the equivalent of~\cref{eq:correlator_par}:
$\smash{\bar S_{hij}^\l(k,x) = \sum_{ee'}{\cal F}_{\evec{k}hij}^{see'}(k_0,x) P_{\vec{k}hij}^{ee'}}$, one immediately finds $\big(k_0 - \bar\omega_{{\bm k} ij}^{ee'}) {\cal F}_{\vec{k}hij} ^{see'} = 0$. This is a spectral equation whose solutions are distributions
\begin{equation}
{\cal F}_{\vec{k} h ij} ^{see'}(k_0,x) = 2\pi f_{\vec{k} h ij}^{see'}(t,\evec{x}) \delta(k_0 - \bar\omega_{{\bm k} ij}^{ee'}),
\label{eq:spectral_F}
\end{equation}
where $f_{\vec{k} h ij}^{see'}$ are some shell-functions that parametrize the correlation functions $S^s_{il}(k,x)$. The cQPA shell solutions~\cref{eq:spectral_F} were found and used to derive QKEs for fermions and bosons in the spectral limit in~\cite{Her081,Her10,Her11,Fid11,Jukkala:2019slc}. The frequencies $\bar\omega_{{\bm k} ij}^{ee'}$ are given by
\begin{equation}
\bar\omega_{{\bm k} ij}^{ee'} \equiv \sfrac{1}{2}(\omega^e_{{\bm k} i} + \omega^{e'}_{{\bm k} hj}).
\label{eq:cQPA_shells}
\end{equation}
We display these shells for two-neutrino mixing in figure~\cref{fig:shell structures}. A solution corresponding to a spectral shell $\smash{k_0=\bar\omega_{{\bm k} ij}^{ee'}}$ has the oscillation frequency $\smash{2\Delta\omega_{{\bm k} ij}^{ee'}}$, given by~\cref{eq:shell-frequencies}. The particle-antiparticle coherence solutions with $e\neq e'$ reside at $k_0\approx 0$ and oscillate very rapidly in comparison to flavour coherence solutions with $e=e'$ but $i\neq j$ that form tight bundles with the usual mass-shell solutions with $e=e'$ and $i=j$. We will use the vast difference in the oscillation frequencies between the particle-antiparticle and flavour mixing to derive a much simpler evolution equation limited to only  flavour mixing in section~\cref{subsec:particle-antiparticle} below.

%
%

\paragraph{Homogeneous Ansatz with the spectral limit}

The Ansatz~\cref{eq:asa1} is a more general construction than the spectral cQPA solution, but it reduces to~\cref{eq:spectral_F} when one uses spectral free theory solutions for the pole functions. The free spectral function in the Wigner space is given by:
\begin{equation}
\phantom{Ha}
\bar {\mathcal{A}}_{ij}(k,x) = \pi \textrm{sgn}(k_0)(\slashed{k}+m_i)\gamma^0\delta(k^2-m_i^2) \delta_{ij}
= \pi\delta_{ij} \sum_a P^a_{\evec{k}i}\delta(k_0-\omega a_{\evec{k}i}).
\label{eq:free_spec_fun}
\end{equation}
Here it is simpler to use the direct space representation of this function, which is just the time-evolution operator:
\begin{equation}
2\bar {\cal A}(u_0,v_0) = e^{i{\cal H}_{{\bm k}}(u_0-v_0)}.
\label{eq:time_evol_op}
\end{equation}
Using~\cref{eq:correlator_par} to write the local correlator appearing in~\cref{eq:asa1} in the projective representation. Then, employing~\cref{eq:time_evol_op} with the rules given below equation~\cref{eq:projection ope}, we immediately get
\begin{equation}
\begin{split}
  \delta \bar S_{\evec{kx}ij}^s (t; u_0, v_0) 
   & = \sum_{habln} 
   2\bar {\cal A}_{\vec{kx}il}(u_0,t) 
               \delta f_{\vec{k} h ln}^{sab}(t,\evec{x}) P^{ab}_{\evec{k}hln} 
   2\bar{\cal A}_{\vec{kx} nj}(t,v_0),
\\
   &= \sum_{hab} \frac{1}{2\bar\omega_{\evec{k}ij}^{ab}}
                 \delta \hat f_{\vec{k} h i j}^{sab}(t,\evec{x}) D_{\vec{k} h i j}^{ab} \gamma^0
                 \exp(-i\omega^a_{\evec{k}i} u_0\!+\!i\omega^b_{\evec{k}j} v_0),
\end{split}
\label{eq:asa2}
\end{equation}
where $s = <,>$ and $\delta \hat f_{\vec{k} h i j}^{sab} \equiv {\rm exp} (2i\Delta\omega_{\evec{k}ij}^{ab} t)\delta f_{\vec{k} h i j}^{sab}$ and we defined
\begin{equation}
D_{\vec{k} h i j}^{ab}
  \equiv 2\bar\omega^{ab}_{\evec{k}ij} P^{ab}_{\evec{k}hij}\gamma^0 
     =   ab \hat N^{ab}_{\evec{k}ij} 
         P_{\evec{k}h}(\slashed k_{i}^a+m_i)(\slashed k_{j}^b+m_j),
\label{eq:D_definition}
\end{equation}
with $(\smash{k_{i}^a)^\mu \equiv (\omega^a_{\evec{k}i},\evec{k})}$ and $\smash{\hat N_{\evec{k}ij}^{ab} \equiv N_{\evec{k}ij}^{ab}\bar\omega^{ab}_{\evec{k}ij}/(2\omega^a_{\evec{k}i} \omega^b_{\evec{k}j})}$. Moving to the Wigner space in frequency and assuming $t = \sfrac{1}{2}(u_0+v_0)$, equation~\cref{eq:asa2} becomes (now written for $\delta S^s_{ij}$ without the bar)
\begin{equation}
i\delta S_{ij}^s (k,x) 
     = 2\pi\sum_{hab} \delta f_{\vec{k} h i j}^{sab}(t,\evec{x})  
            \frac{1}{2\bar\omega_{\evec{k}ij}^{ab}} 
            D^{ab}_{\evec{k}hij} \delta(k_0 - \bar{\omega}_{\vec{k} ij}^{ab}).
  \label{eq:cQPA_prop_Wigner}
\end{equation}
This is the just the spectral cQPA-result~\cref{eq:spectral_F}. We shall see in section~\cref{sec:collision_integrals} that the non-vanishing phase factors for $t \neq \frac{1}{2}(u_0+v_0)$ have a crucial role in ensuring correct 4-momentum conservation over the internal vertices in loops contributing to collision integrals~\cite{KaiPa23}. 

In the diagonal limit $a=b$ and $i=j$ the structure $\smash{D_{\vec{k} h i j}^{ab}}$ reduces to the standard form: $\smash{D_{\vec{k} h i i}^{aa} = P_{\evec{k}h}(\slashed k_{i}^a+m_i)}$. We then get: 
\begin{equation}
i\delta S_{ii}^{saa}(k,x) 
     = 2\pi\sum_{h} \delta f_{\vec{k}hii}^{saa}(t,\evec{x})  
           P_{h\evec{k}}(\slashed k_{i}^a+m_i) \frac{1}{2\omega^a_{{\bm k}i}}\delta(k_0 - \omega^a_{\evec{k}i}).
\label{eq:mass-shell-solution}
\end{equation}
This solution has the same form as the most general spectral adiabatic background solution depending on flavour, helicity and frequency, proving that in the spectral limit the background solutions can indeed be merged into diagonal transient solutions. One can then write the full solution in the same form as~\cref{eq:cQPA_prop_Wigner_full}:
\begin{equation}
i S_{ij}^s (k,x) 
     = 2\pi\sum_{hab} f_{\vec{k} h i j}^{sab}(t,\evec{x})  
            \frac{1}{2\bar\omega_{\evec{k}ij}^{ab}} 
            D^{ab}_{\evec{k}hij} \delta(k_0 - \bar{\omega}_{\vec{k} ij}^{ab}).
\label{eq:cQPA_prop_Wigner_full}
\end{equation}
\vskip-0.1truecm \noindent
where $f_{\vec{k} h i j}^{sab} = \delta_{ab}\delta_{ij}f_{{\rm eq},\bm {k}ii}^{sa} + \delta f_{\vec{k} h i j}^{sab}$.

We stress that the spectral limit for adiabatic solutions is convenient and often sufficient, but not obligatory assumption for the analysis of~\cref{eq:AH4} and the self-energy terms that appear in the equation. One could add non-trivial gradient corrections and finite widths to~\cref{eq:AH4} and to spectral function~\cref{eq:free_spec_fun}, although this would come with a considerable amount of additional tedium. Our goal has been to balance between the full generality and the simplicity of notation for the benefit of both approaches.

%
\subsection{Hermitian self-energy corrections}
\label{sec:sub_spectral_adiabatic}
%

In essentially all problems in neutrino physics one can neglect dispersive and finite width corrections to background solutions and in the definition of the projective basis. However, we {\em have} included forward scattering corrections to~\cref{eq:AH4}. As we discussed at the end of the section~\cref{sec:decoupling}, we can use different Hermitian self-energy functions in the pole- and the background equations and in equation~\cref{eq:AH4}. This freedom induces forward scattering corrections to~\cref{eq:AH4} even when the pole equations are treated in the free theory limit. 

Moreover, we can {\em evaluate} the effective Hermitian self-energies~\cref{eq:sigma_eff} using the vacuum spectral function~\cref{eq:free_spec_fun}. This immediately reduces the effective self-energies into simple products:
\begin{equation}
\bar \Sigma^{\rm H}_{{\rm eff},\evec{k}ij}(t,\evec{x}) 
= (\bar \Sigma_{\rm H} * 2\bar {\cal A})_{\evec{kx},ij}(t,t)
 = \sum_a \nolimits \bar\Sigma_{{\rm H},\evec{kx}ij}(a\omega_{\evec{k}j})P^a_{\evec{k}j}.
\label{eq:sigma_eff2}
\end{equation}
The forward scattering coefficients then become
\begin{equation}
(\W^{{\rm H}ee'}_{\evec{k}hij})^{l}_{a}
=\Tr[P_{\vec{k}hji}^{e'e}\,\bar{\Sigma}_{{\rm H}\vec{k}il}(a\omega_{\evec{k}l})
     P_{\vec{k} hlj}^{ae'}]. 
\label{eq:WrmH}
\end{equation}
This expression involves ordinary self-energy function instead of the effective one. One should note that the self-energy function inherits its energy sign from the nearest energy projector to the right from the self-energy function.

\begin{figure}
    \centering
    \includegraphics[width=0.85 \textwidth]{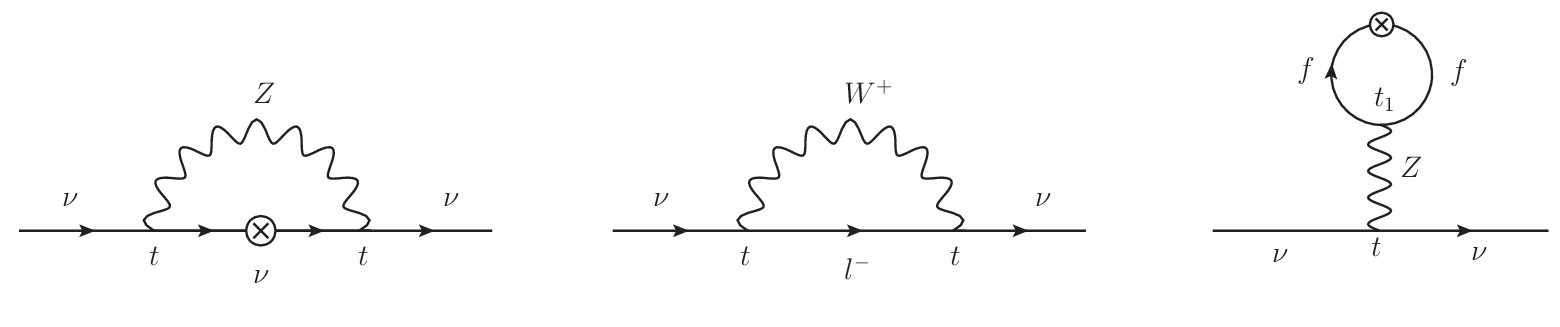}
    \caption{One-loop diagrams which contribute to light neutrino self-energy.}
\label{fig:1loop self-energy}
\end{figure}

In general $\bar{\Sigma}_{{\rm H}\evec{k}}$ can depend on the dynamical quantities we are set out to solve. An example is given by the first diagram in the figure~\cref{fig:1loop self-energy} which contains an internal neutrino line. This correction can cause a strong back-reaction from local neutrino densities, which have been shown give rise to interesting new phenomena in the early universe~\cite{Enqvist:1999zs,Kainulainen:2001cb,Hannestad:2015tea,Hannestad:2013pha,Abazajian:2012ys} as well as in the core collapse supernovae and in the accretion discs around compact objects~\cite{Volpe:2023met}. Often $\Sigma_{{\rm H}\evec{k}}$ is dominated by the interactions with the background however, and in such cases one can use some adiabatic (thermal of finite density) approximation for ${\Sigma}_{{\rm H}\vec{k}}$. We give a simpler treatment for such cases below and postpone a      direct evaluation of the diagrams involving neutrinos to section~\cref{sec:General_self-energy}.

\paragraph{Simple 1-loop weak gauge corrections.}

%
%

Any homogeneous self-energy function can be expressed in terms of structures~\cref{eq:basis1} and evaluated in the projective basis using table~\cref{tab:reduction_formulae}. The weak gauge interactions in particular induce a one-loop self-energy for light neutrinos corresponding to diagrams shown in fig.~\cref{fig:1loop self-energy}. This self-energy can be written in the following general form:
\begin{equation}
\begin{split}
    \bar \Sigma_{{\rm H},ij}(k,x) & = \gamma^0 ( a_{ij}\slashed{k} + b_{ij} \slashed{u})  P_L \\
    & = \bigl(k_0 a_{ij} + b_{ij}\bigr) P_L - a_{ij} \evec{\alpha} \cdot \evec{k} P_L, \\
    & \rightarrow \Big((k_0 + h|\evec{k}|) a_{ij} + b_{ij}\Big) P_L \equiv V_{\vec{k} h ij}(k_0,x) P_L,
    \label{eq:self-energy exact form}
    \end{split}
\end{equation}
where $a_{ij}(k,x)$ and $b_{ij}(k,x)$ are some space-time varying flavor matrices and $u^\mu$ is the plasma 4-velocity. In the second line we went to the plasma rest frame $u^\mu = (1,\vec{0})$ and in the third line we used the fact that operators $\evec{\alpha}\cdot \hat{\evec{k}}P_L$ and $-hP_L$ have the same projective representation in the vacuum eigenbasis. In this case the forward scattering coefficient tensor becomes 
\begin{equation}
\label{eq:forward scattering}
(\W^{ e'e}_{\evec{k}hij})^{l}_{a}
= \sigma_{\vec{k} h lji}^{eae'} V_{\vec{k} hil}(a\omega_{\evec{k}l},x),
\\
\end{equation}
where $V_{\vec{k} h ij}(k_0,x)$ is the matter potential experienced by propagating neutrinos, defined above in~\cref{eq:self-energy exact form} and the tensor $\sigma_{\vec{k} lji}^{abc}\equiv {\rm Tr}[P^{ca}_{\evec{k}hil}P_LP^{bc}_{\evec{k}hji}]$ is given by (see~\cref{eq:basis1}):
\begin{equation}
    \sigma_{\evec{k} h lji}^{a b c}  \equiv \frac{1}{2}N_{\evec{k}il}^{ca} N_{\evec{k}ji}^{bc}
    \Big( \frac{\hat{P}_{\vec{k} h l}^{a}}{(N_{\vec{k} ji}^{bc})^{2}} 
        + \frac{\hat{P}_{\vec{k} h j}^{b}}{(N_{\vec{k}il}^{ca})^{2}} 
      - \hat{P}_{\vec{k}hi}^{c}\Big(\frac{1}{(N_{\vec{k} lj}^{ab})^{2}} 
            - ab \frac{m_l m_j}{\omega_{\bm{k}l} \omega_{\bm{k}j}} \Big) \Big),
    \label{eq:sigma_tensor}
\end{equation}
where $P_{\evec{k}hi}^a \equiv \frac{1}{2}(1-ahv_{\evec{k}i})$ is yet another useful projector function. In the ultra relativistic (UR) limit $P_{\evec{k}hi}^a \rightarrow \delta_{a,-h}$. Even with thermal assumption, the expression~\cref{eq:sigma_tensor} is much more general than the ones presented in the literature so far, see \eg~\cite{Serreau:2014cfa}. It will simplify significantly in the frequency diagonal and in the UR-limits to be considered below.

\paragraph{Renormalization.}

Let us briefly discuss the role of renormalization in our quantum transport formalism. The key point is that renormalization only affects the vacuum parts of the correlation functions arising from a need to regulate and properly define the vacuum self-energy corrections. From the results of section~\cref{sec:decoupling} it is then clear that renormalization only affects the vacuum limit of the pole and the background statistical functions.  Moreover, the SD-equations for these functions decouple in the spectral limit, after which renormalization procedure only concerns the Hermitian self-energy function and can be done using the standard field theory methods, augmented to account for the flavour mixing. We refer the reader for example to~\cite{Juk21} for a recent explanation of the procedure. For our current purposes, we can assume that our Hermitian self-energy functions are properly renormalized and we are using renormalized masses and couplings to parametrize our theory.

%
\section{Limiting cases and extensions}
\label{sec:limits}
%

So far our results are very general and in particular valid for arbitrary masses. This results in complex tensor structures in the projected equations and in the projected self-energies. In some problems, such as the production of flavour mixing particles from background fields, this complexity is unavoidable. Considerable simplifications arise if one can neglect the particle-antiparticle mixing and even more if one can take the UR-limit. Indeed, if one was only interested in the flavour oscillations, the particle-antiparticle mixing is but an unnecessary complication. We now show how to remove these structures from~\cref{eq:AH4} by a simple integration~\cite{Juk21}.

%
\subsection{Integrating out the particle-antiparticle oscillations}
\label{subsec:particle-antiparticle}
%

As was already pointed out, the particle-antiparticle oscillations have very high frequencies and they should average out in the usually much slower flavour oscillation scale. Indeed from equations~\cref{eq:AH4} and~\cref{eq:shell-frequencies} the leading time-dependence of the particle-antiparticle coherence functions is $f_{\vec{k}hij}^{e -e}(t) \sim \exp(-2ie \bar{\omega}_{\vec{k}ij}t)$ with $2\bar\omega_{\evec{k}ij} = \omega_{\evec{k}i}+\omega_{\evec{k}j}$, while for the flavour mixing functions $f_{\vec{k}hij}^{ee}(t) \sim \exp(-2ie \Delta \omega_{\evec{k}ij}t)$ with $2\Delta \omega_{\evec{k}ij} = \omega_{\evec{k}i}-\omega_{\evec{k}j}$. This suggests~\cite{Juk21} to take a Weierstrass transform of equation~\cref{eq:AH4}:
\begin{equation}
    \int \dd t' \, W(t,t') \bigl[\text{e.o.m.}(t')\bigr]
    \text{,}
\end{equation}
where $W(t,t') \sim \exp(-(t-t')^2)/2\sigma^2$. Given hierarchy $\Delta  \omega_\evec{k}\ll \bar  \omega_\evec{k}$, we can choose a $\sigma$ such that $1/\Delta \omega_{\vec{k}} \gg \sigma \gg 1/\bar{\omega}_{\vec{k}}$. Then all terms in the equation which vary in the flavor scale are essentially unchanged by the transform, while the terms proportional to the coherence functions get exponentially suppressed:
\begin{equation}
    \int \dd t' \, W(t,t') c^e_{\vec{k}}(t')\delta f^{e -e}_{\vec{k} h}(t')
    \sim c^e_{\vec{k}}(t)\delta f^{e -e}_{\vec{k} h}(t) \exp\bigl(-2(\bar\omega_{\vec{k}} \sigma)^2\bigr),
\end{equation}
where $c^e_{\vec{k} h}$ stands for any coefficient of $\delta f^{e -e}_{\vec{k} h}$ in the equation of motion which is assumed to vary only in the flavour scale. The Weierstrass transform thus induces a {\em coarse-graining} with a temporal resolution scale $\sigma$ which effectively washes out the particle-antiparticle mixing from the master equation.

In the absence of the particle-antiparticle mixing the tensor structures in~\cref{eq:AH4} simplify considerably. In particular the velocity tensor in the spatial gradient term becomes just
\begin{equation}
({\cal V}_{\bm{k} h ij}^{e'e})_{aa'} \rightarrow \frac{e}{2} \delta_{ee'}\delta_{aa'}\delta_{ae} (v_{\evec{k}i} + v_{\evec{k}j}) \equiv e\delta_{ee'}\delta_{aa'}\delta_{ae} {\bar{v}}_{{\bm k}ij},
\end{equation}
where $\bar{v}_{{\bm k}ij}$ is the average velocity of the flavour states $i$ and $j$ with momentum ${\bm k}$. The projected density matrix equation restricted either to particle or antiparticle sector then becomes:
\begin{equation}
\boxed{
\begin{aligned}
\partial_t  f_{\vec{k} h ij}^{e} + \bar{v}_{{\bm k}ij}\, \hat{\bm k} \cdot \bm{\nabla} f_{\vec{k} h ij}^{e} = & -2ie\Delta\omega_{\bm{k} ij}  f_{\vec{k} h ij}^{e} 
+ {\rm Tr}[ \bar{\mathcal{C}}^\l_{{\rm H},\vec{k} hij}   P_{\vec{k} h ji}^{ee}] 
\\
 & + f_{\vec{k} h  il}^{e} i\sigma_{\vec{k} h jil}^{eee} V_{\vec{k} hlj}(e\omega_{\evec{k}l})
   - i\sigma_{\vec{k} h lji}^{eee} V_{\vec{k} hil}(e\omega_{\evec{k}l}) f_{\vec{k} h lj}^{e}.
\label{eq:AH5}
\end{aligned}
}
\end{equation}
No sum over the energy signs remains here, but the sum over the repeated flavour index $l$ persists. The tensor $\smash{\sigma_{\vec{k} h lji}^{eee}}$ still has a complex mass-dependence for general kinematics. We continue to defer the evaluation of the collision integral to a later stage. Here we also made explicitly the Feynman-Stueckelberg interpretation and replaced everywhere:
\begin{equation}
\evec{k} \rightarrow e\evec{k} \qquad {\rm and \;\; (then)} \qquad f^{\l ee}_{(e\evec{k})hij} \rightarrow ef^{e}_{\evec{k}hij}.
\end{equation}
After this identification $f^+_{\evec{k}h}$ refers to particle and $f^-_{\evec{k}h}$ to antiparticle flavour density matrix. Equation~\cref{eq:AH5} is relevant for studying for example the resonant leptogenesis. It presents a generalization from~\cite{Juk21} in that it gives explicit expressions for the dispersive corrections that so far have not been included in any leptogenesis calculation. Because~\cref{eq:AH5} also allows for (adiabatic) evolution in the spatial coordinate, it could be easily used to accurately model the heavy neutrino production and the associated lepton number violating processes in the collider experiments with arbitrary heavy state kinematics. 

%
\subsection{The ultra-relativistic limit}
\label{sec:UR-limit}
%

Equation~\cref{eq:AH5} simplifies further in the ultra-relativistic limit. UR-limit can be always taken when dealing with light neutrinos and it is therefore useful to show the equation explicitly in this case. Using UR-expansion for energies it is easy to show that 
\begin{equation}
\label{eq:normalization_fac_ur}
N_{\evec{k}ij}^{ab} \approx \delta_{a,-b} + \delta_{a,b} \frac{2|\evec{k}|}{m_i+m_j}
\quad \Rightarrow \quad
 \sigma_{\vec{k} h ilj}^{eee} \approx \delta_{e,-h}.
\end{equation}
With this result we immediately get:
\begin{equation}
\begin{split}
\sigma_{\vec{k} h lji}^{eee} V_{\vec{k} hil}(e\omega_{\evec{k}l},x)
&\approx \delta_{e,-h}V_{\evec{k}hil}(e|\evec{k}|,x) 
 \equiv (V^e_{\evec{k}h})_{il}\\
\sigma_{\vec{k} h jil}^{eee} V_{\vec{k} hlj}(e\omega_{\evec{k}l},x) 
&\approx \delta_{e,-h}V_{\evec{k}hlj}(e|\evec{k}|,x) 
\equiv (V^e_{\evec{k}h})_{lj}.
\end{split}
\end{equation}
The forward scattering terms now collapsed to the familiar light neutrino matter potentials. If we further define an effective matter Hamiltonian:
\begin{equation}
(H^e_{\evec{k}h})_{ij} = e \delta_{ij}\omega_{\evec{k}i} + (V^e_{\evec{k}h})_{ij},
\label{eq:matter_hamiltonian}
\end{equation}
and diagonal velocity matrix $v_{{\bm k}ij} \equiv \delta_{ij}|\evec{k}|/\omega_{\evec{k}i}$, we can write equation~\cref{eq:AH5} in the UR-limit in the compact, familiar form of a density matrix evolution equation:
\begin{equation}
\boxed{
\partial_t  f_{\vec{k} h}^{e} 
           + \sfrac{1}{2} \{ v_{\bm k}, \hat{\bm{k}} \cdot \bm{\nabla} f_{\vec{k} h}^{e} \} =
- i[H_{\evec{k}h}^e, f_{\vec{k} h}^{e}] + \bar{\mathcal{C}}_{\vec{k} h}^{e},\strut}
\label{eq:AH2_UR}
\end{equation}
where $\smash{(\bar{\mathcal{C}}_{\vec{k} h}^{e})_{ij} \equiv {\rm Tr}[ \bar{\mathcal{C}}^\l_{{\rm H},\vec{k}hij} P_{\vec{k} h ji}^{ee}]}$. When dealing with light neutrinos over relatively small propagation distances, one can further set $v_{{\bm k}ij} \rightarrow \delta_{ij}$, in which case the spatial flow term reduces to 
$\smash{ \sfrac{1}{2} \{ v_{\bm k}, \hat{\bm{k}} \cdot \bm{\nabla} f_{\bm{k} h}^{e} \} 
\rightarrow \hat{\evec{k}}\cdot \bm{\nabla}f_{\bm{k} h}^{e}}$. 

Equation~\cref{eq:AH2_UR} looks deceivingly simple, but it still contains all information of flavour coherences and forward scattering potentials in the UR-limit. It also has the same form as early UR-limit kinetic equations derived by the S-matrix or operator formalism techniques~\cite{Enqvist:1990ad,Enqvist:1990ek,Enqvist:1991qj,Sigl:1992fn} and it is sufficient for almost all light-neutrino physics applications, from laboratory experiments to light neutrino interactions in the early universe and within high density astrophysical objects. Before we turn to crucial issue of computation of the collision integrals we still discuss two issues related to the matter Hamiltonian and the choice of the basis one uses for the pole funcctions. 

\paragraph{Flavour and the matter eigenbases}

So far we have worked exclusively in the vacuum basis. However, given the effective Hamiltonian, we can perform rotations to the flavour or matter bases in the usual manner. In particular the transformation between the flavor and mass basis is just a constant rotation. 
\begin{equation}
H_{\vec{k}h}^{\textrm{fl},e} \equiv U H_{\vec{k}h}^{e} U^{\dagger},
\end{equation}
where in the standard case with three light neutrinos $U$ would be the usual PMNS mixing matrix. A similar transformation could be performed to go to the matter basis where $H^e_{\evec{k}h}$ is diagonal. The matter basis, which always depends on ${\evec k}$ and possibly on $h$, in general also rotates along the neutrino path, because $V^e_{\evec{k}h}=V^e_{\evec{k}h}(x)$. As a result $U_m \equiv U^e_{\evec{k}h}(x)$ depends on the space-time coordinate $x$ as well, which gives rise to the additional Liouville terms after the rotation into the matter basis: 
$\partial_t f \rightarrow \partial_t f_m  + [U_m (\partial_t U_m^\dagger), f_m]$ and 
$\{ v_{\bm k}   \, , \, \hat{\bm{k}} \cdot \bm{\nabla} f \} \rightarrow 
 \{ v_{{\bm k}m}\, , \, \hat{\bm{k}} \cdot \bm{\nabla} f_m + [U_m (\hat{\bm {k}}\cdot\bm{\nabla} U_m^\dagger),f_m ]\}$, where $v_{{\bm k}m} = U_m v_{{\bm k}m}U_m^\dagger$.

%
\subsection{Quasiparticle basis}
\label{sec:quasiparticle_basis}
%

Until now we have used vacuum solutions both for the projective representation and in the reduction of the effective self-energy functions. This is a very good assumption in all light neutrino physics applications and in the leptogenesis problem. However, in some cases dispersive corrections can change the phase space structure significantly. A well known example is the infrared region of a thermal plasma with gauge interactions, where new collective hole excitations appear~\cite{Weldon:1982bn,Weldon:1989bg,Weldon:1989ys}. This situation is realized in some electroweak baryogenesis scenarios~\cite{Farrar:1993hn,Cline:2012hg,Cline:2013gha,Cline:2017qpe} due to strong flavour blind QCD interactions. For illustration we briefly consider this case where additional weak flavour mixing interactions can be treated perturbatively as discussed above. 

The QCD-interactions are vector-like and induce a Hermitian self-energy correction, which in the hard thermal loop  (HTL) approximation has the form
\begin{equation}
\Sigma_\H^{\rm th} = a \slashed {k} + b \slashed{u}.
\end{equation}
Because $\Sigma_\H^{\rm th}$ is flavour diagonal, we can work in the vacuum mass eigenbasis. Neglecting for the moment all other corrections, the inverse propagator for the system can be written as
$S^{-1} = rnp_0\gamma^0 - r\evec{\gamma} \cdot \evec{k} - m_i$, where $r \equiv 1-a$ and $rnk_0 \equiv rk_0 - b$. This propagator has two branches of poles given by:
\begin{equation}
  nrk_0  = \pm \sqrt{|r{\evec{k}}|^2 + m_i^2} 
 \quad \Rightarrow \quad k_0 = \omega^{\rm pl}_{i\pm}(|\evec{k}|,T).
\label{eq:QBEwith-dispersion}
\end{equation}
The positive sign corresponds to particle and the negative sign to the hole solutions of~\cite{Weldon:1989bg,Weldon:1989ys}. One can now derive the effective Hamiltonian near these quasiparticle shells~\cite{Kai21}:
\begin{equation}
{\cal H}^\pm_{{\evec{k}h}} = \delta_{ij}\frac{1}{n_i}\big(\evec{\alpha}\cdot \evec{k} + \frac{m_i}{r_i}\gamma^0 \big)\Big|_{\abs{k_0}=\omega_{i\pm}^{\rm pl}}.
\label{eq:matter_hamiltnioan_resummed}
\end{equation}
Thermal corrections are manifested mainly via the nontrivial refractive index $n_i$. One can proceed to construct the energy projectors using~\cref{eq:matter_hamiltnioan_resummed} in entirely analogous manner to the vacuum case. One should also include the thermal wave-function corrections for the quasistates, see \eg~\cite{Kai21}. Similarly, one can use quasi-particle generalization of the spectral function in the evaluation of the various self-energy functions. Finally the perturbative flavour changing interactions can be added on top of this structure similarly to the vacuum case.

An even simpler generalization to the energy-shell projectors, relevant for the leptogenesis problem, is to replace the constant masses by a time dependent masses. In this case the energy projectors pick up a time dependence through masses that gives rise to additional terms proportional to $\partial_t m_i$ in the projected equations~\cref{eq:AH4,eq:AH5,{eq:AH2_UR}}. For explicit expressions see~\cite{Juk21}.

%
\section{Collision integrals}
\label{sec:collision_integrals}
%

In this section we show how to compute the collision integrals with coherent states in our transport equations. Our approach is different from section~\cref{sec:sub_spectral_adiabatic} where we assumed generic structures for the Hermitian self-energy function and worked out their projections. Here we construct the expansion of the collision integrals directly in the projected basis. That is we will compute the collision integral traces:
\begin{equation}
\Tr[ \bar{\mathcal{C}}^\l_{{\rm H},\vec{k}hij} P_{\vec{k} h ji}^{e' e}]
= \sum_{ab}\bar{\mathcal{C}}^{\l ab}_{{\rm H},\vec{k}hij} \Tr[P_{\vec{k} h ij}^{a b}  P_{\vec{k} h ji}^{e' e}]
= \bar{\mathcal{C}}^{\l ee'}_{{\rm H},\vec{k}hij},
\label{eq:coll_integral_components}
\end{equation}
where in the last step we used the orthogonality of the energy projection operators and the normalization~\cref{eq:normalization_fac}. We now insert the expression~\cref{eq:master_coll} for $\bar{\mathcal{C}}^\l_{{\rm H},\vec{k}hij}$ into~\cref{eq:coll_integral_components} and proceed similarly to what we did with the Hermitian self-energy function terms in section~\cref{sec:projected_master}. We can then write the generic collision integral as follows
\begin{equation}
\bar{\mathcal{C}}^{\l ee'}_{{\rm H},\vec{k}hij} = 
\frac{1}{2}\sum_{l,a}\nolimits 
 \Big[ (\W^{{\g}ee'}_{\evec{k}hij})^l_a   f_{\vec{k} h lj}^{\l a e '}
     + [(\W^{\g e'e}_{\evec{k}hji})^l_a]^*f_{\vec{k} h il}^{\l e a }  
     - ( > \leftrightarrow < )\Big],
\label{eq:collision_integral_generic_form}
\end{equation}
where $(\W^{s e'e}_{\evec{k}hji})^l_a$-tensors are defined similarly to $(\W^{{\rm H} e'e}_{\evec{k}hji})^l_a$-tensor in~\cref{eq:WrmH}: 
\begin{equation}
 (\W^{s ee'}_{\evec{k}hij})^l_a =\frac{1}{4\bar\omega_{\evec{k}ij}^{ee'}\bar\omega_{\evec{k}lj}^{ae'}}
       {\rm Tr}[\,\Sigma^{s a}_{\evec{k} il}\, D_{\evec{k}hlj}^{ae'} 
       \gamma^0 D_{\evec{k}hji}^{e'e}], 
\label{eq:full_collision_integral_term}
\end{equation}
for $s = >,<$. Here we preferred to wrote the projection operators in the $D$-tensor notation of equation~\cref{eq:D_definition}. The key quantity in the expression~\cref{eq:full_collision_integral_term} is the self-energy function:
\begin{equation}
\Sigma^{s a}_{\bm{k} il} \equiv \int {\rm d} w_0 \Sigma^{s}_{\bm{k} il} (t,w_0) 
 \exp(i\omega^a_{{\bm k}l}(t-w_0))
 = \Sigma_{{\rm out},{\bm k}hil}^s(\omega^a_{{\bm k}l}),
\label{eq:sigma_a}
\end{equation}
where again $s =>,<$. An essential element in the following analysis is the fact that the homogeneous Ansatz~\cref{eq:asa1}, which reduced temporal convolutions to simple products, also reduces an arbitrary function $\Sigma^{sa}_{\bm{k} il}$, with any number of internal lines, computable in the local limit. From the reduction process we can infer simple rules for a diagrammatic construction of collision integrals. We give the main points of the derivation and the final results here. More details can be found in the companion paper~\cite{KaiPa23}.

%
\subsection{General reduction process}
%

Any diagram contributing to the self energy~\cref{eq:sigma_a} consists of a number of vertices connected by propagators, with an integration over the spacetime coordinates in each internal vertex. Wigner transforming propagators then introduces a momentum integral for each internal line and a group of phase factors that after integration give rise to 4-momentum conservation over vertices. For the spatial coordinates and momenta this works out in the usual manner, but the time coordinate requires more attention. Moreover, the internal lines are divided to statistical (cut) and pole propagators, following the standard rules of the thermal field theory~\cite{Bellac:2011kqa,Fid11,KaiPa23}. Each cut propagator introduces a statistical $f$-factor that gets associated with an external state in the interaction process, while the pole propagators give rise to the internal resonances or generate loop corrections to collision rates. 

It is simplest to work in the 2-time representation and we continue to assume the free theory limit for the pole functions. In this case the dynamical and the background solutions can be combined as explained in section~\cref{sec:spectral_limit}. Following~\cref{eq:cQPA_prop_Wigner} the full statistical correlation function can then be written as:
\begin{equation}
  \bar S_{\evec{kx}ij}^s (t; u_0, v_0) 
   = \sum_{hab} \frac{1}{2\bar\omega_{\evec{k}ij}^{ab}}
                 f_{\vec{k} h i j}^{sab}(t,\evec{x}) D_{\vec{k} h i j}^{ab}  
                 \exp(-2i\Delta^{ab}_{{\bm k}hij}t-i\omega^a_{\evec{k}i} u_0  
                                                  +i\omega^b_{\evec{k}j} v_0),
\label{eq:Wightman_gen}
\end{equation}
where $s = <,>$. In this article we consider only the gauge interactions and treat the gauge fields as non-coherent resonances. This means that we only need the 2-time representation of the standard gauge-field propagator:
\begin{equation}
{\cal D}_{\mu\nu}({\bm q};u_0,v_0) = \int \frac{{\rm d}q_0}{2\pi} {\cal D}_{\mu\nu}(q) e^{-iq_0(u_0-v_0)},
\label{eq:gauge_propagator}
\end{equation}
where ${\cal D}_{\mu\nu}(q)$ is the usual momentum space propagator function. Extension to gauge fields, or additional scalar fields, that are a part of the non-equilibrium quantum plasma is straightforward, but to keep discussion simple we do not present it explicitly here.

%
\begin{figure}[t]
\centering
\includegraphics[width=.7\textwidth]{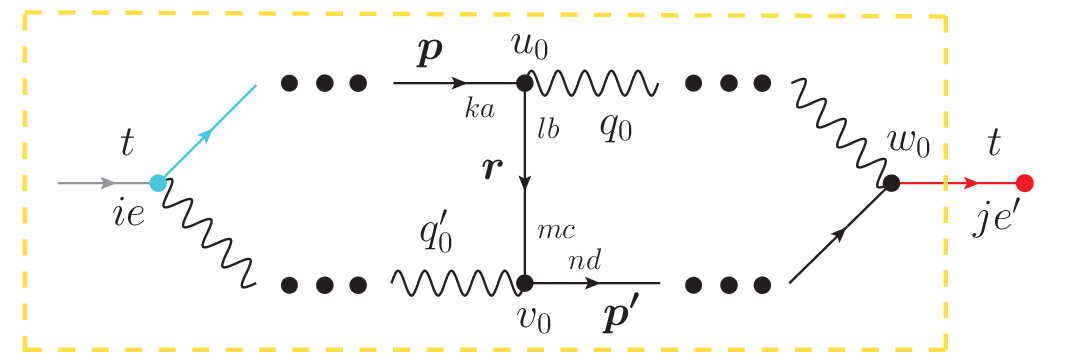}
\caption{A section of a generic diagram contributing to the projected local collision integral~\cref{eq:coll_integral_components}. The dashed outline shows the extent of the self-energy contribution and the detail at the center displays insertion of nontrivial coherent propagators~\cref{eq:Wightman_gen}.}
\label{fig:genericloop}
\end{figure}
%

Figure~\cref{fig:genericloop} shows the propagator~\cref{eq:Wightman_gen} associated with the internal time coordinates $u_0$ and $v_0$ in a generic self-energy diagram contributing to the self energy~\cref{eq:sigma_a}. Remarkably, the phase factors in~\cref{eq:Wightman_gen} appear with {\em different} frequencies at coordinates $u_0$ and $v_0$ when either $e\neq e'$ or $i\neq j$. This difference is crucial to ensure the correct energy conservation at both vertices even when the connecting propagator has support on a constant energy shell $\bar\omega^{ee'}_{\evec{k}ij}$. Including also the phase factors from the gauge-field propagators~\cref{eq:gauge_propagator}, we see that the integral over $u_0$ results in $2\pi\delta(a\omega_{\evec{p}k}-b\omega_{\evec{r}l}-q_0)$ and the integral over $v_0$ gives $2\pi\delta(c\omega_{\evec{r}m}-d\omega_{\evec{p}'n}+q^\prime_0)$. 
As usual, the delta-functions from vertices kill all frequency integrals associated with the propagators, leaving one extra delta-function that gives a generalized energy conservation for the process in question~\cite{KaiPa23}. Using the delta-functions one can also show that the sum of all explicit $t$-dependent phases, which appear in the definition~\cref{eq:sigma_a} and in the cut propagators~\cref{eq:Wightman_gen}, vanishes exactly for any process~\cite{KaiPa23}. Instead of giving general proofs of these statements, we will show below how this works in  specific examples.

Having gotten rid of all phase factors we see that each internal cut propagator effectively contributes the following factor to the diagram:
\begin{equation}
\begin{split}
       &\sum_{hab} \int \! \frac{{\rm d}^3\evec{k}}{(2\pi)^3 2\bar\omega_{\evec{k}ij}^{ab}} 
                      f_{\vec{k} h i j}^{s ab}(t,\evec{x}) D_{\vec{k} h i j}^{ab}
\\ 
     = &\int \frac{{\rm d}^4 k}{(2\pi)^4} 
                     \Big[ 2 \pi\sum_{hab} \frac{1}{2\bar\omega_{\evec{k}ij}^{ab}} 
                      f_{\vec{k} h i j}^{s ab}(t,\evec{x}) D_{\vec{k} h i j}^{ab} 
                      \delta(k_0-\bar\omega_{\evec{k}ij}^{ab}) \Big]  
     \equiv \sum_{hab}\int \frac{{\rm d}^4 k}{(2\pi)^4} iS^{sab}_{hij}(k,x).
\end{split}  
\label{eq:twotime_red}
\end{equation}
The differential fraction in the integral in the first line in~\cref{eq:twotime_red} defines a natural phase space density factor and $\smash{f_{\vec{k} h i j}^{sab}(t,\evec{x})}$ is the corresponding phase space distribution function. This leaves $\smash{D_{\vec{k} hij}^{ab}}$ as the sole contribution to the scattering matrix element. Alternatively, one can use the singular cQPA-form given on the second line as the Feynman rule for the cut propagators with the associated full four dimensional momentum integral. 

\begin{figure}[t]
\centering
\includegraphics[width=.75\textwidth]{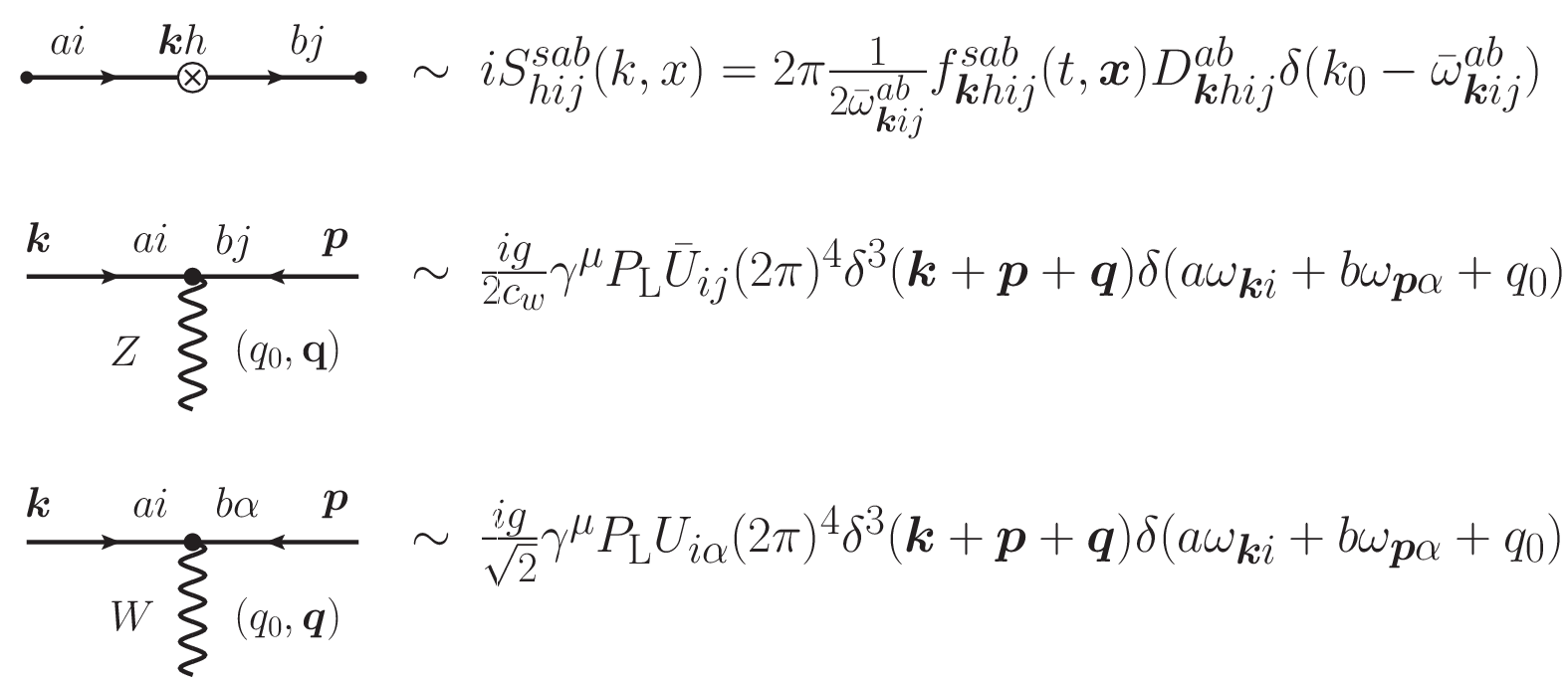}
\caption{The Feynman rules for the internal lines and weak interaction vertices associated with coherent neutrino propagators. In the $W$-boson vertex $U_{i\alpha}$ reduces to the PMNS-matrix in the case of pure active-active mixing ($\alpha$ refers to lepton flavour). In the $Z$-boson vertex $c_w = \cos\theta_w$ and the mixing matrix $\bar U_{ij}$ reduces to $\mathbbm{1}$ for pure active-active mixing. The rules generalize to arbitrary Lorentz and flavour structures in an obvious way.}
\label{fig:feynman_fules1}
\end{figure}

These results can be summarized by the Feynman rules for evaluating self-energy diagrams shown in figure~\cref{fig:feynman_fules1}. In addition to these rules one should associate each propagator line with an integral over the four-momentum and sums over the helicity, the energy-sign and the flavour indices. For the pole propagators one should use just the usual momentum space Feynman rules.

%
%

%
\subsection{Collision integral from two-loop gauge diagrams}
\label{sec:collision_two_loop}
%

\begin{figure}[t]
\centering
\begin{subfigure}{.28\textwidth}
  \includegraphics[width=1.0\textwidth]{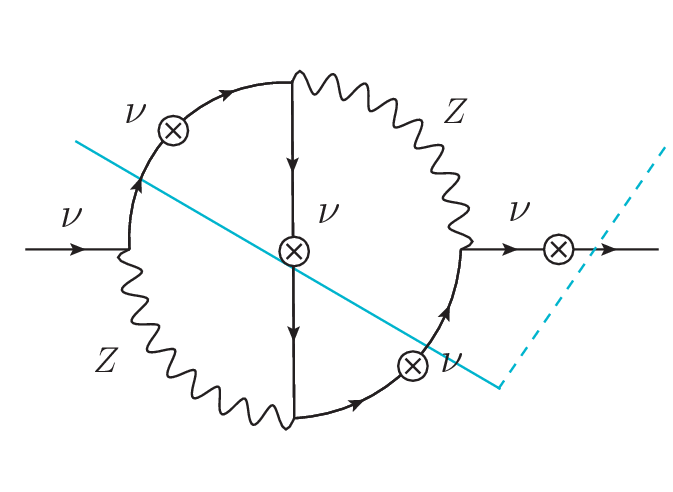}
\end{subfigure} 
\begin{subfigure}{.28\textwidth}
  \includegraphics[width=1.0\textwidth]{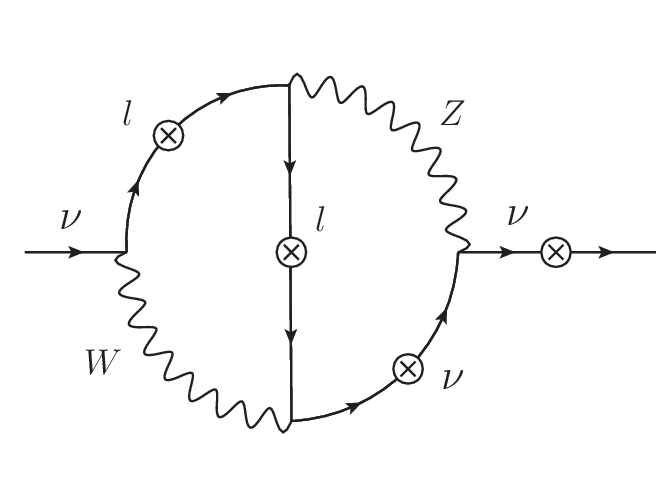}
\end{subfigure}
\begin{subfigure}{.28\textwidth}
  \centering
\includegraphics[width=1.0\textwidth]{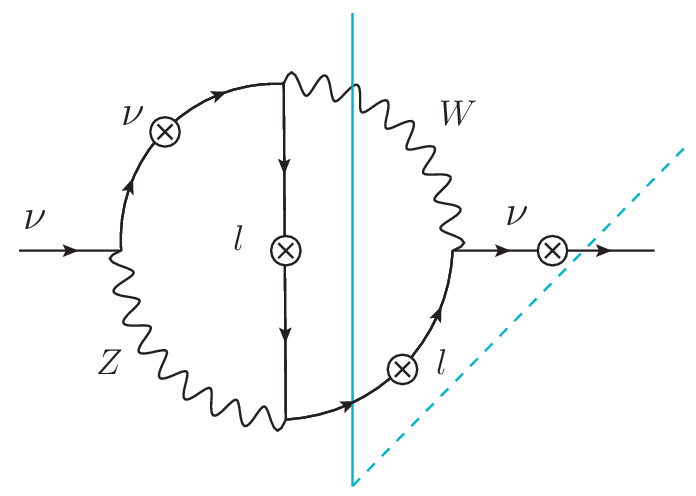}
\end{subfigure}
\begin{subfigure}{.28\textwidth}
    \includegraphics[width=1.0\textwidth]{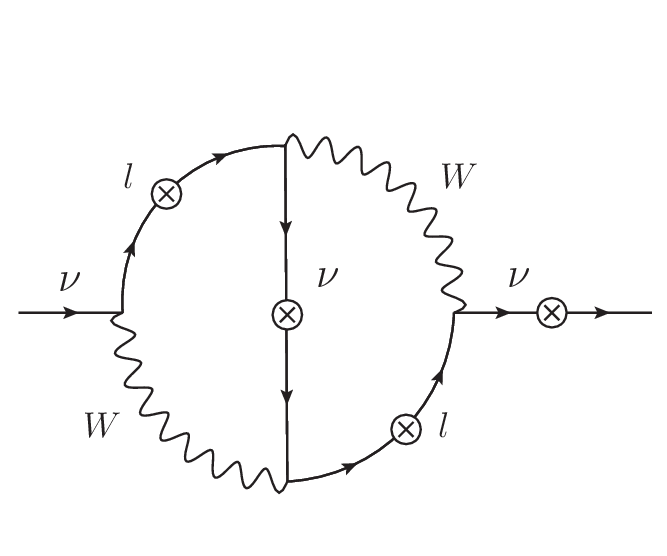}
\end{subfigure}
\begin{subfigure}{.28\textwidth}
  \includegraphics[width=1.0\textwidth]{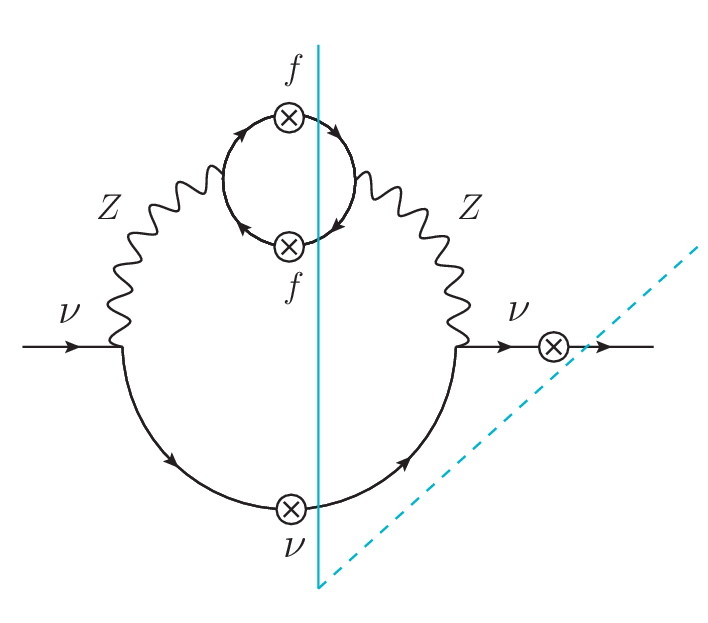}
\end{subfigure} 
\begin{subfigure}{.28\textwidth}
  \includegraphics[width=1.0\textwidth]{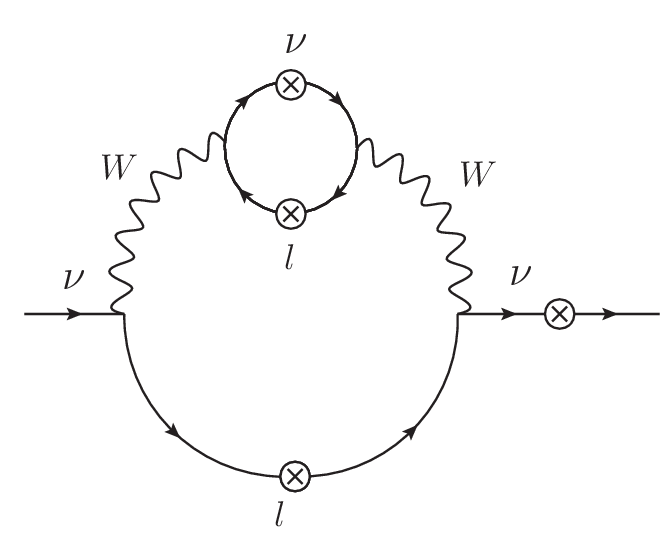}
\end{subfigure}
\caption{The 2-loop Feynman diagrams for the self-energy function that give rise to 2-2 neutrino scattering processes are shown. Crosses in propagator lines denote the possibly coherent propagator function~\cref{eq:twotime_red}, and the light blue lines are thermal cuts which are explained in the text.}
\label{fig:2loop_graphs}
\end{figure}

In figure~\cref{fig:2loop_graphs} we show all two-loop self-energy diagrams that give rise to the 2-2 scattering terms in collision integrals mediated by the weak gauge interactions. Only the four first diagrams are two-particle irreducible (2PI). The last two diagrams on the second row should not be included in the full SD approach, where they would be accounted for by the one-loop corrections to the gauge-boson equations. However, when gauge bosons are treated as non-dynamical resonances, the 2PI-hierarchy is partially broken and these diagrams must be added as perturbative corrections to gauge boson propagators. They eventually produce the squared matrix elements for the $s$, $t$ and $u$-channels for a given scattering process while the 2PI-diagrams give rise to the interference terms between the different channels. 

In addition to scattering terms these self-energies create a large number of 1-3 decay and inverse decay terms as well as vertex corrections to 1-2 decays. Different processes correspond to different cuts on the self-energy diagrams following standard rules of the finite temperature field theory~\cite{Bellac:2011kqa,Fid11,KaiPa23}. For illustration we show cuts that give rise to $Z$-mediated 2-2 scatterings and 1-3 decays between neutrinos and other fermions in the first and fifth diagrams and a cut producing $W$-boson 1-2 decay correction in the third diagram. The 2-2 scatterings and the 1-3 decay corrections are further distinguished by the frequency sign signatures in the overall momentum conservation function following the standard kinematic analysis which we shall elaborate more in~\cite{KaiPa23}.

%
%

\paragraph{The self energy function $\smash{\Sigma^{\l a}_{{\rm ZZ},il}}$}

\begin{figure}[t]
\centering
    \includegraphics[width=1.0\textwidth]{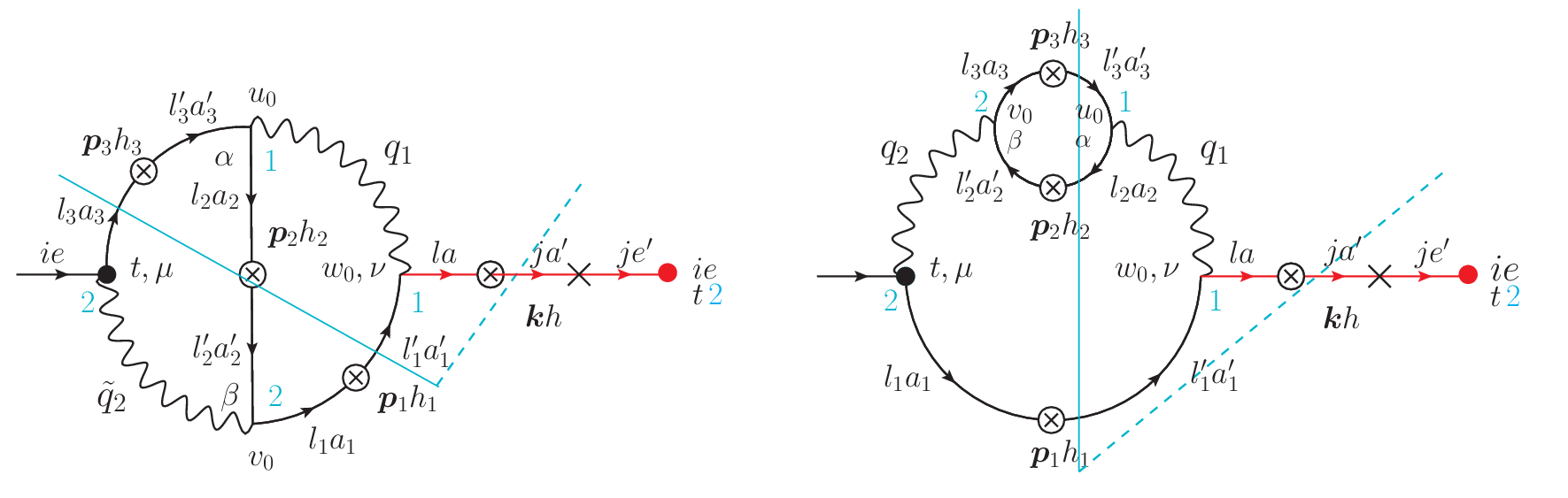}
\caption{Two-loop graphs that contribute to the matrix element squared of neutrino-neutrino scattering through s and t channels are shown with explicit index structures and cuts. The black dot implies the starting point of the evaluation of the trace. The red propagator is the dependent momentum propagator, see section~\cref{subsec:simplified_Feynman_rules_1} for explanation.}
\label{fig:ZZ combined}
\end{figure}

We now work out the self-energy function $\Sigma^{\l a}_{{\rm ZZ},il}$ that describes scatterings between coherent neutrinos mediated by $Z$-gauge bosons. The relevant diagrams are the first in the upper and the second in the lower row of fig.~\cref{fig:2loop_graphs}. We show these diagrams again in figure~\cref{fig:ZZ combined} including labels for the momenta and the discrete indices for the intermediate states. To alleviate the complexity of notation we will be using the shorthand
\begin{equation}
S_{h_il_il_i'}^{sa_ia_i'}(k_i,x) \equiv S^s_{{\rm X}_i}(k_i,x).
\end{equation}
Note that the rightmost propagators marked red in figure~\cref{fig:ZZ combined} are not part of the self-energy function, but they do contribute to the collision integral according to~\cref{eq:full_collision_integral_term}. The light blue numbers at each vertex correspond to the Keldysh time-path indices~\cite{Bellac:2011kqa,KaiPa23}. They allow reading which type of propagator corresponds to each line: $S^{12}_{X_i}=S^\l_{X_i}$ and $S^{21}_{X_i}=S^\g_{X_i}$ for the statistical cut propagators and $i{\cal D}^{11}_{\mu\nu} = i{\cal D}_{\mu\nu}$ and $i{\cal D}^{22}_{\mu\nu} = (i{\cal D}_{\mu\nu})^*$ for the standard gauge field propagators. The cut-line thus passes through all statistical propagators in the diagram. Using these instructions and the Feynman rules given in figure~\cref{fig:feynman_fules1} it is easy to see that the interference diagram gives: 
\begin{align}
        i\Sigma^{\l a,{\rm int}}_{{\rm ZZ}, il} (k,x)  = & \Big(\frac{ig}{2c_w}\Big)^4 
         \sum_{\{{\rm X}_i\}} \nolimits \bar U_{il_3} \bar U_{l_3'l_2} \bar U_{l_2'l_1} \bar U_{l_1'l} 
         \int_{\{p_i,q_i\}} i{\cal D}_{Z \alpha\nu}(q_1) (i{\cal D}_{Z \mu\beta}(\tilde q_2))^*
        \nonumber \\
        &  \times 
        (2\pi)^9 \delta^3({\bm q}_1 - {\bm p_3} + {\bm p}_2)
                 \delta^3({\bm q}_1 - {\bm k}   + {\bm p}_1)
                 \delta^3(\tilde {\bm q}_2 - {\bm p}_1 + {\bm p}_2) 
        \nonumber \\ \phantom{\int_{\{p_i\}}}
        &  \times 
        (2\pi)^3 \delta(q_{10} - \omega^{a'_3}_{{\bm p}_3l'_3} + \omega^{a_2}_{{\bm p}_2l_2})
                 \delta(q_{10} - \omega^{a}_{{\bm k}l}         + \omega^{a'_1}_{{\bm p}_1l'_1})
                 \delta(\tilde q_{20} - \omega^{a_1}_{{\bm p}_1l_1}   + \omega^{a'_2}_{{\bm p}_2l'_2}) 
        \nonumber \\ 
        &  \times 
          \gamma^{\mu}    P_{\rm L} \; i S_{X_3}^\l(p_3,x) 
        \;\gamma^{\alpha} P_{\rm L} \; i S_{X_2}^\g(p_2,x)
        \;\gamma^{\beta}    P_{\rm L} \; i S_{X_1}^\l(p_1,x) 
        \;\gamma^{\nu}  P_{\rm L},
    \label{eq:sigmaZZ1a}
 \end{align}
where $\smash{\int_p \equiv \int {\rm d}^4p/(2\pi )^4}$ and the curly brackets indicate groups of indices to be summed or variables to be integrated over. It is easy to perform the $q_1$ and $q_2$-integrals using the delta-functions from vertices, which leaves out the one overall momentum conserving delta function. After performing the integrations one finds:
\begin{equation}
        i\Sigma_{{\rm ZZ}, il}^{<a,{\rm dir}} (k,x) =  \sum_{{\{\rm X}_i\}} \nolimits
         \int\dd{{\rm PS}_3} 
         A^{{\rm int},a}_{\evec{k} il \{\evec{p}_i, {\rm X}_i\}} \; \Lambda^{\l}_{\{\evec{p}_i, {\rm X}_i\}}(x),
    \label{eq:sigmaZZ1b}
\end{equation}
where the phase space integral is defined as:
\begin{equation}
    \int \dd {{\rm PS}_3} \equiv \int 
      \Big[ \prod_{i=1,3} \frac{{\rm d}^3\evec{p}_i}{(2\pi)^3 2\bar\omega_{{\rm X}_i}}\; \Big] 
        (2\pi)^4 \delta^4(k^a_l - p_{1l_1'}^{a_1'}
                                + p_{2l_2}^{a_2} -p_{3l_3'}^{a_3'}),
\label{eq:phase_space_standard}
\end{equation}
where the energy sign indices tell whether a given term contributes to a 2-2 scattering or a 1-3 decay channel. The distribution functions in~\cref{eq:sigmaZZ1a} were gathered into the factor
\begin{equation}
\Lambda^{\l}_{\{\evec{p}_i, {\rm X}_i\}}(x) \equiv 
      f^{\l}_{{\rm X}_1\evec{p}_1}(x)
      f^{\g}_{{\rm X}_2\evec{p}_2}(x)
      f^{\l}_{{\rm X}_3\evec{p}_3}(x),
\label{eq:distribution_lambda_element}      
\end{equation}
and the part that eventually contributes to the matrix element was defined as:
\begin{align}
A^{{\rm int},a}_{\evec{k}il\{\evec{p}_i,{\rm X}_i\}} = 
\Big(\frac{ig}{2c_w}\Big)^4 & \; \bar U_{il_3} \bar U_{l_3'l_2} \bar U_{l_2'l_1} \bar U_{l_1'l}
       {\cal D}_{Z\alpha\nu}(q_1) {\cal D}^{*}_{Z\mu\beta}(\tilde q_2) 
\nonumber \\ \times &
        \gamma^{\mu}    P_{\rm L} \, D_{{\rm X}_1\evec{p_1}}
        \gamma^{\alpha} P_{\rm L} \, D_{{\rm X}_2\evec{p_2}}
        \gamma^{\beta}  P_{\rm L} \, D_{{\rm X}_3\evec{p_3}}
        \gamma^{\nu}    P_{\rm L},
\label{eq:A_int}
\end{align}
where the gauge-boson 4-momenta are 
$q_1 = (\omega^a_{\evec{k}l}\! - \omega^{a_1'}_{\evec{p}_1l_1'}; \evec{k} \!-\evec{p}_1)$ and  $\tilde q_2 = (\omega^{a_1}_{\evec{p}_1l_1}\! - \omega^{a_2'}_{\evec{p}_2l_2'};\evec{k} \!-\evec{p}_3)$.

%
%
%

Evaluation of the "direct" diagram shown in the right in figure~\cref{fig:ZZ combined} proceeds analogously. The phase space-integral~\cref{eq:phase_space_standard} and the element~\cref{eq:distribution_lambda_element} containing the distribution functions are the same as in the interference diagram. The only difference stems from the different way of connecting the fermion lines and from the value of the gauge boson momentum $q_2$, which give rise to a different $A$-factor:
\begin{align}
A^{{\rm dir},a}_{\evec{k}il \{\evec{p}_i,{\rm X}_i\}} = 
-\Big(\frac{ig}{2c_w}\Big)^4 & \; \bar U_{il_1} \bar U_{l_1'l} \bar U_{l_2'l_3} \bar U_{l_3'l_2}
       {\cal D}_{Z\alpha\nu}(q_1) {\cal D}^{*}_{Z\mu\beta}(q_2) 
\nonumber \\ \times &
        \, \gamma^{\mu}     P_{\rm L} \, D_{{\rm X}_1\evec{p_1}}
        \gamma^{\nu}  P_{\rm L} \, 
        \Tr\big[\gamma^{\alpha} P_{\rm L} D_{{\rm X}_2\evec{p_2}} \gamma^{\beta}    P_{\rm L} 
                  D_{{\rm X}_3\evec{p_3}}\big],
\label{eq:A_direct}
\end{align}
where $q_1 = (\omega^a_{\evec{k}l}\! - \omega^{a_1'}_{\evec{p}_1l_1'}; \evec{k} \!-\evec{p}_1)$ and $q_2 = (\omega^{a_3}_{\evec{p}_3l_3}\! - \omega^{a_2'}_{\evec{p}_2l_2'};\evec{k} \!-\evec{p}_1)$.
\vskip0.1truecm

Note that both 3-momenta in the direct channel~\cref{eq:A_direct} are the same ${\bm q}_1={\bm q}_2={\bm k}-{\bm p}_1$, while in the interference channel $\tilde {\bm q}_2={\bm k}-{\bm p}_3$ is different. This is as expected, since direct diagram accounts for the $s$- and $t$-channels, while interference diagram accounts for their interference. Identifying different channels by Mandelstam variables is not obvious in presence of particle-antiparticle mixing and was done above based on 3-momenta. In the frequency diagonal limit the situation is simple. For example the $t$- and $u$-channel particle-particle collisions correspond to $a=e=a_i=a_i'=1$ for all $i$, while the $t$- and $u$-channel particle-antiparticle collision correspond to $a=e=a_3=a_3'=1$ and  $a_1=a_1'=a_2=a_2'=-1$. These are just examples from the different processes embedded in the generic collision integral. Identifying and classifying all different channels is straightforward but somewhat tedious task of standard kinematic analysis. For more details see~\cite{KaiPa23}.

\paragraph{Collision integral for Z-mediated processes}

We have now collected all the pieces needed to write down the full collision integral corresponding diagrams in figure~\cref{fig:ZZ combined}. Inserting the self-energies computed above to equation~\cref{eq:coll_integral_components} and doing some simple reorganization we find
\begin{equation}
\overline{\mathcal{C}}_{{\rm ZZ, H},\vec{k} h ij}^{<ee'} 
    =  \sum_{\rm Y} \nolimits
         \frac{1}{2\bar\omega_{\bm{k}lj}^{aa'}} \int\dd{\rm{PS}_3} 
         \Big[ \sfrac{1}{2}(\mathcal{M}^2)_{\bm{k}hij \{\evec{p}_i,\rm Y\}}^{ee'} 
         \Lambda_{\bm{k} hj\{\bm{p}_i,\rm Y\}}(x) + \;  (h.c.)^{e'e}_{ji} \Big], 
\label{eq:collZZ1}
\end{equation}
where we collected all summed indices into curly brackets ${\rm Y} = \{ {\rm X}_i, a, a', l\}$. The phase space factor is the same as in~\cref{eq:phase_space_standard} and the factor containing all distribution functions is
\begin{equation}
\Lambda^{\l}_{\evec{k} hj\{\evec{p}_i,\rm Y\}}(x) = 
     f^{\l aa'}_{\evec{k} hlj}(x) \,
     f^{\g}_{{\rm X}_1\evec{p}_1}(x) \,
     f^{\l}_{{\rm X}_2\evec{p}_2}(x)\,
     f^{\g}_{{\rm X}_3\evec{p}_3}(x)\, - ( \, > \leftrightarrow < \,).
\label{eq:lambda factor}
\end{equation}
Finally, the effective matrix element squared is defined as
\begin{equation}
(\mathcal{M}^2)_{\evec{k} h i j \{\evec{p}_i,{\rm Y}\}}^{ee'} \equiv
      \frac{1}{2\bar\omega_{\evec{k}ij}^{ee'}}
         \Tr[A^a_{\evec{k}il\{\evec{p_i} {\rm X_i}\}} D_{\evec{k} h' l j}^{aa'} \gamma^0 D_{\evec{k} h j i}^{e'e}],
\label{eq:effective_matrix_element}
\end{equation}
with 
\begin{equation}
A^a_{\evec{k}il\{\evec{p_i} {\rm X_i}\}} \equiv A^{{\rm int},a}_{\evec{k}il\{\evec{p_i} {\rm X_i}\}} + A^{{\rm dir},a}_{\evec{k}il\{\evec{p_i} {\rm X_i}\}},
\end{equation}
where the quantities in the right hand side are given in equations~\cref{eq:A_int} and~\cref{eq:A_direct}. Remember that 
$f^{\g ab}_{{\bm p}hij} = a\delta_{ij}\delta_{ab}-f^{\g ab}_{{\bm p}hij}$ in general and the rule $\smash{\bar f^{\l,\g}_{\evec{k}hij} = - f^{\g,\l --}_{(-\evec{k})hij}}$ for translation between negative frequency and antiparticle distributions. Note that, as emphasized by our notation, the frequency and flavour indices get flipped in the Hermitian conjugate term in~\cref{eq:effective_matrix_element}, as is clear from~\cref{eq:collision_integral_generic_form}.

Let us stress that in the presence of mixing the collision integral components are in general complex; indeed, we found already in section~\cref{sec:spectral_limit} that a function $f^{ab}_{{\bm k}hij}(t,{\bm x})$ has a leading phase $-2\Delta \omega^{ab}_{{\bm k}hij}t$. From~\cref{eq:lambda factor} and the phase space constraint in~\cref{eq:phase_space_standard} it is then easy to work out that the leading phase of the phase factor term $\Lambda$ (and hence that of the whole integrand) is
\begin{equation}
\varphi_\Lambda = (\omega^{e'}_{{\bm k}j} - \omega^{a_1}_{{\bm p}_1l_1} 
           + \omega^{a_2'}_{{\bm p}_2l_2'} - \omega^{a_3}_{{\bm p}_3l_3})t.
\end{equation}
This means that the integrand is a rapidly oscillating function in the particle-antiparticle mixing scale in all cases except the fully diagonal limit, where $e=e'$ and $a_i=a_i'$ for all $i$ and in the fully anti-diagonal limit, where $e=-e'$ and $a_i=-a_i'$. This implies that also the collision integral averages to zero under the Weierstrass coarse-graining of the equation of motion in the limit discussed in section~\cref{subsec:particle-antiparticle}. 

Despite the compact notation, expression~\cref{eq:collZZ1} is a very complex object that encompasses all flavour and particle-antiparticle mixing effects for arbitrary neutrino masses and kinematics. It still displays only the familiar structures from the usual Boltzmann theory and all complexity is reduced to a bookkeeping of the indices labeling the states. In particular, the phase-space factor and the dependence on the distribution functions are universal features independent of the structure of the interactions. The interaction matrix element is more complex than the usual result for non-mixing states, but its evaluation is formally straightforward and can be easily done using symbolical routines. We will evaluate~\cref{eq:effective_matrix_element} in the UR-limit in section~\cref{sec:UR-limit_results} below. For now, we comment that the simple, intuitive form of our collision integral, which clearly separates the different flavour and frequency structures, appears to be in stark contrast with other existing QKE computations in the literature~\cite{Vlasenko:2013fja,Blaschke:2016xxt}.

%
\begin{figure}[t]
\centering
\includegraphics[width=0.9\textwidth]{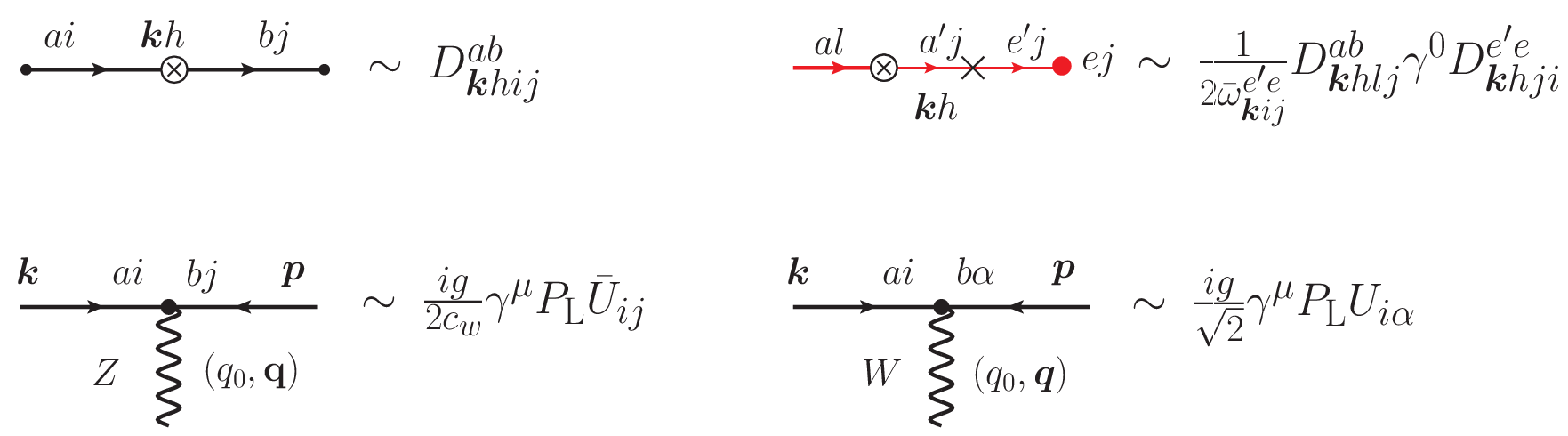}
\caption{Simplified Feynman rules for computing the squared matrix element directly. The first propagator function is to be used in all internal lines and the special DMP propagator (see text for discussion), shown by red color applies for the outgoing line in the diagram. For the definition of quantities in the vertex factors see figure~\cref{fig:feynman_fules1}.}
\label{fig:feynman_rules2}
\end{figure}
%

%
\subsection{Simple Feynman rules for the matrix element squared}
\label{subsec:simplified_Feynman_rules_1}
%

The simple factorization of the collision integral to universal phase space elements and a dynamical matrix element squared allows us to deduce a very simple set of rules to evaluate the collision integrals directly without the need to repeat the integration procedure for each new diagram and set of interactions. The rules we now spell out should be obvious from the above derivation.
\begin{itemize}
\item{ Draw the loop diagrams that contribute to a given interaction process to the desired order in perturbation theory, and assign unique momentum variable and flavor and frequency indices for each internal propagator line in the graph, allowed by the interaction vertices.}
\item{ Assign the Keldysh-path indices to all vertices to isolate the cut that gives rise to the desired interaction process. You only need to evaluate $\Sigma^\g= \Sigma^{21}$ directly, so the first index is always 2 and the last 1.} 

\item{Read off the phase space functions contributing to the $\Lambda$-factor from all internal cut propagator lines. Add the external phase space factor $f_{\evec{k} hlj}^{\l aa'}/2\omega^{aa'}_{\evec{k}hlj}$, which is associated with the external, dependent momentum propagator (DMP), marked red in diagrams in figure~\cref{fig:ZZ combined}.}

\item{ Determine the phase space density factor with the overall energy conserving delta function. This depends on the number of loops in the diagram and the cut one is interested in. At two loops with a cut leading to 2-2 scatterings this was simple: we isolated the internal 11-propagator and eliminated its frequency and momentum from one of its end-vertex delta-functions using the other one. }
\item{ Compute the matrix element squared using the Feynman rules shown in fig.~\cref{fig:feynman_rules2}. Start from the equivalent of the black dot in the diagrams in figure~\cref{fig:ZZ combined} and follow the direction of momentum in the graph. For each internal cut-line insert the standard propagator shown in the first diagram in~\cref{fig:feynman_rules2}. Add the DMP at the end of the fermion line it is connected to. For each ("22") "11" line use the (anti) Feynman propagator. Add the DMP at the end of the fermion line it is connected to. Take a trace over the Dirac indices.}
\item{ Divide the result by two and add the Hermitian conjugate, accounting for the index changes as indicated in~\cref{eq:effective_matrix_element}.}
\end{itemize}

We invite the reader to verify that these rules indeed allow writing~\cref{eq:collZZ1,eq:lambda factor,eq:effective_matrix_element} directly from the diagrams shown in figure~\cref{fig:ZZ combined}. These rules can also be extended, in an obvious way, to any other interaction types. One particularly interesting application is the Leptogenesis problem where the gauge interactions are replaced by scalar Yukawa interactions. In that case one also encounters Majorana neutrinos which require some special rules that can be found for example in~\cite{Juk21}.

%
\subsection{Explicit results in the UR-limit}
\label{sec:UR-limit_results}
%

We return now to evaluate the effective matrix element~\cref{eq:effective_matrix_element} explicitly in the UR-limit. This limit is interesting for practical applications and because it allows to make a direct contact with some known results in the literature. We begin by pointing out some general simplifications: first the orthogonality of the energy projectors immediately sets $a'=e'$ independent of the type of interactions. Second, in the case of neutrino-neutrino interactions the trace calculation simplifies due to the orthogonality of the chirality operators which immediately gives:
\begin{equation}
    \begin{split}
       P_{\rm L} D_{\evec{k} h ij}^{ab} P_{\rm R} = ab\hat{N}_{\evec{k}ij}^{ab} P_{\rm L} P_{\evec{k} h} (m_j \slashed{k}_i^a + m_i \slashed{k}_j^b)P_{\rm R} 
       \approx \delta_{a,-h}  \delta_{a,b} P_{\rm L}\slashed{p}^a P_{\rm R},
     \end{split}
\label{eq:helicity_simplification}
\end{equation}
where the last result applies in the UR-limit. To arrive at this result we used the expansion
$\smash{\hat N^{ab}_{\evec{k}ij} \approx \delta_{a,-b}/(2|\bm{k}|) + \delta_{a,b}/(m_i+m_j)}$. Similarly, one can show that 
\begin{equation}
P_{\rm L} D_{\evec{k} h' l j}^{ae'} \gamma^0 D_{\evec{k} h j i}^{e'e}P_{\rm R} \approx \delta_{a,e}\delta_{e,e'}\delta_{e,-h} 2e|\bm{k}| P_{\rm L} \slashed{k}^eP_{\rm R},
\label{eq:helicity_simplification_2}
\end{equation}
in the UR-limit.  From now on we will also assume that the $Z$-boson is very heavy in the energy scale of interest, which means that we can set: $\smash{{\cal D}_{Z\mu\nu} = g_{\mu\nu}/M_{\rm Z}^2}$. With these simplifications the matrix element~\cref{eq:effective_matrix_element} becomes simple to evaluate. The result is:
\begin{equation}
\begin{split}
(\mathcal{M}^2)_{\evec{k}hij \{\evec{p}_i,{\rm Y}\}}^{ee'} 
= - 32G_{\rm F}^2 \delta_{e,e'}\delta_{a,e}\delta_{e,-h} 
                & \delta_{a_1,a_1'}\delta_{a_1,-h_1} 
                  \delta_{a_2,a_2'}\delta_{a_2,-h_2} 
                  \delta_{a_3,a_3'}\delta_{a_3,-h_3} 
   \\ &\; \times 
       \Big( \bar U^{\rm 4,dir}_{ilX} + \bar U^{\rm 4,int}_{ilX} \Big) \; 
       (k^e \cdot p_2^{a_2})(p_1^{a_1} \cdot p_3^{a_3}),
\label{eq:matrixeleKron}
\end{split}
\end{equation}
where $G_{\rm F}$ is the Fermi constant and $\bar U^{\rm 4,dir}_{ilX} \equiv \bar U_{il_1} \bar U_{l_1'l} \bar U_{l_2'l_3} \bar U_{l_3'l_2}$ and $\bar U^{\rm 4,int}_{ilX} \equiv \bar U_{il_3} \bar U_{l_3'l_2} \bar U_{l_2'l_1} \bar U_{l_1'l}$.  Note that in the UR-limit all collision terms with particle-antiparticle mixing drop out at the leading order. The mixing terms would appear as ${\cal O}(m/E)$-corrections only, which we have dropped above. Such terms also would have a large mixing phase.

\paragraph{Standard Model (SM) limit}

To make equations even more transparent, we now assume the SM-limit where the neutral current rotation matrices are trivial: $\bar U_{ij} = \delta_{i,j}$. Making use of the large number of Kronecker delta functions in~\cref{eq:matrixeleKron}, one can show that the general collision term (\ref{eq:collZZ1}) now reduces simply to
\begin{equation}
    \begin{split}
       \overline{\mathcal{C}}_{{\rm ZZ, H},\vec{k} h ij}^{< e e}  =
       - 16 G_{\rm F}^2  \, \delta_{e,e'}&  \delta_{e,-h} 
        \sum_{\{a_i\}} \frac{1}{2|\vec{k}|} \int {\rm d PS}_3 
        \; (k^e \cdot p_{2}^{a_2})(p_{1}^{a_1} \cdot p_{3}^{a_3})
        \Lambda^e_{\{a_i\}ij}[f],
       \end{split} 
    \label{eq:collZZ1_UR1xx}
\end{equation}
where the phase space factor is defined as before, but now the flavour indices are no longer necessary in the kinematic factors:
\begin{equation}
    \int \dd{\rm{PS}_3} = \int 
      \Big[ \prod_{i=1,3} \frac{{\rm d}^3\evec{p}_i}{(2\pi)^3 |2{\evec{p}_i}|}\; \Big] 
        (2\pi)^4 \delta^4(k^e - p_1^{a_1} + p_2^{a_2} - p_3^{a_3}),
\label{eq:phase_space_UR}
\end{equation}
and all flavour dependence is in the phase space factor:
\begin{equation}
\begin{split}
 \Lambda^e_{\{a_i\}ij}[f] = 
 \sum_{l,l',l''} \Big\{ \Big(& f^{\g a_3a_3}_{ {\bm p}_3 -a_3 l'' l'  }
                               f^{\l a_2a_2}_{ {\bm p}_2 -a_2 l'  l'' }
                               f^{\g a_1a_1}_{ {\bm p}_1 -a_1 i   l   } 
                               f^{\l ee}_{{\bm k}-elj}
                       \\ 
                          +  & f^{\g a_3a_3}_{ {\bm p}_3 -a_3 i   l'  }
                               f^{\l a_2a_2}_{ {\bm p}_2 -a_2 l'  l'' }
                               f^{\g a_1a_1}_{ {\bm p}_1 -a_1 l'' l   } 
                               f^{\l ee}_{{\bm k}-elj}   \Big)
                         -  ( > \leftrightarrow < ) \Big\} + h.c.\, .
\label{eq:phasefactor_again}
\end{split}
\end{equation}
The first combination of $f$-functions in~\cref{eq:phasefactor_again} arises from the direct diagram and the second set from the interference diagram. Note that while pure neutral current processes do not give rise to source terms in~\cref{eq:AH4}, nontrivial source and therefore nontrivial mixing contributions would be produced by charged currents. Equations~\cref{eq:collZZ1_UR1xx,eq:phasefactor_again} give the correct neutral current collision integral for such a setup.

\paragraph{Neutrino-antineutrino scattering}

Next consider the special case of $\nu\bar\nu-\nu\bar\nu$-scattering in the UR and SM-limits. To get this process we must set\footnote{
%
%
The indices 1 and 3 correspond to dummy variables and are interchangeable here.} 
%
%
$e=1 = a_3 = 1$ and $a_1 = a_2 = -1$ in~\cref{eq:collZZ1_UR1xx,eq:phase_space_UR,eq:phasefactor_again}. To impose the Feynman-Stueckelberg relations we first redefine ${\bm p}_i \rightarrow a_i{\bm p}_i$ and ${\bm k}\rightarrow e{\bm k}$. This immediately sets 
$(k^e \cdot p_{2}^{a_2})(p_1^{a_1} \cdot p_3^{a_3}) \rightarrow 
ea_1a_2a_3(k \cdot p_2)(p_1 \cdot p_3)$ with ordinary dot-products. In the present case moreover $ea_1a_2a_3 = 1$. Similarly, the delta-function in the phase space factor now becomes
$\delta^4(k+p_1-p_2-p_3)$ appropriate for the process we are considering. Finally, we identify \eg~$\smash{f^{\l -}_{-{\bm p}_2-(-1)l'l''} = -\bar f^{\g}_{{\bm p}_2+ij}}$. Working similarly with the other distribution functions, we find that the particle distribution factor becomes:
\begin{equation}
    \begin{split}
     \Lambda_{\vec{k} lj \{ \vec{p}_i, {\rm X}_i\}}^{\nu\bar\nu\rightarrow \nu\bar\nu} 
     = \sum_{l,l',l''} \Big\{ \Big( 
       & \bar f^\l_{\vec{p}_1+il} f^\l_{\vec{k}-lj} 
         \bar f^\g_{\vec{p}_2+l''l'}f^\g_{\vec{p}_3-l'l''}
    \\
     + &f^\g_{\vec{p}_3-il'} \bar f^\g_{\vec{p}_2-l'l''} \,
        \bar f^\l_{\vec{p}_1-l''l} f^\l_{\vec{k}-lj}
     - (>\leftrightarrow <)\Big) + h.c. \Big\} .
    \end{split}
    \label{eq:lambda_diagonal}
\end{equation}
Writing furthermore $\smash{f^{\g}_{{\bm k}_2hij} = \delta_{ij} - f^{\l}_{{\bm p}_2hij}}$ and similarly for the antiparticle distributions, this falls into the familar form from the usual Boltzmann theory, except that the expression still contains all information of the flavour mixing.

If we finally take the flavour diagonal limit, all terms in $\Lambda$ become real. We also see that the second (interference) term in~\cref{eq:lambda_diagonal} gives just one term with $i=l=l''=j$. In the first (direct) term however, we only get $i=l=j$ and the term still contains a sum over one flavour: $l'=l''$. That is, when $l' \neq i$ only the first term survives and gives the collision integral for the usual $\nu_i\bar\nu_l-\nu_i\bar\nu_l$-scattering coming from the single $t$-channel diagram. Indeed, the matrix element for this process is easy to compute and the result $|{\cal M}_t|^2 = 32G_{\rm F}^2 (k\cdot p_2)(p_1\cdot p_3)$ is in perfect agreement with the above results. For $l'= i$ both terms contribute and reproduce the result for $\nu_i\bar\nu_i-\nu_i\bar\nu_i$-scattering obtained from summing the $s$- and $t$-channel terms using the usual field theory methods. 
\vskip0.3truecm

\paragraph{Two flavour active-sterile mixing}

Let us finally consider the case of two flavour mixing between an active ($a$) and a sterile ($s$) neutrino in the UR-limit. We will cast the collision integral for this system into the familiar form in the flavour basis. As before, we label the flavour states by Greek and vacuum basis states by Latin letters. In this case the rotation matrix between the flavour and vacuum basis $U$ and the neutral current mixing matrix $\bar U$ are given by:
\begin{equation}
U_{i\alpha} = \left( \begin{array}{cc}
                   c &-s \\
                   s & c 
                   \end{array}\right),
\qquad
\bar U_{\alpha\beta} = \left( \begin{array}{cc}
                 1 & 0 \\
                 0 & 0 
                   \end{array}\right)
\quad \Rightarrow \quad
\bar U_{ij} = \left( \begin{array}{cc}
                 c^2 & cs \\
                 cs & s^2 
                   \end{array}\right),
\end{equation}
where \eg~$c \equiv \cos\theta$, where $\theta$ is the vacuum mixing angle. Now observe that the rotation matrices $\bar U$ within $\bar U^{\rm 4,dir}_{ilX}$ and $\bar U^{\rm 4,int}_{ilX}$ always sandwich the distribution functions $f^{<,>}_{{\rm X}_i\bm{p}_i}$ in equation~\cref{eq:distribution_lambda_element}. Because the matrix element function~\cref{eq:effective_matrix_element} does not depend on flavour in the UR-limit, the rotation amounts to replacing the vacuum basis matrices $\bar U_{ij}$ by their flavour basis representations $\bar U_{\alpha\beta}$ as well as setting $f^{sab}_{\bm{p}hij} \rightarrow U_{\alpha i} f^{sab}_{\bm{p}hij} U^\dagger_{\beta j} \equiv f^{sab}_{\bm{p}h\alpha\beta}$ everywhere. The final result is
\begin{equation}
C^{ee}_{{\rm ZZ,H}\bm{k}h \alpha\beta} 
      = -\left( \begin{array}{cc}
              \;\;\Gamma^{\g e}_{{\rm ZZ,H}\bm{k}h aa} f^{<e}_{\bm{k}haa} &
              \sfrac{1}{2}\Gamma^{\g e}_{{\rm ZZ,H}\bm{k}h aa}\, f^{<e}_{\bm{k}has} \\
              \sfrac{1}{2}\Gamma^{\g e}_{{\rm ZZ,H}\bm{k}h aa}\, f^{<e}_{\bm{k}hsa}& 0 
        \end{array}\right) - \big( < \leftrightarrow >\big) ,
\end{equation}
where we used $(f^{<e}_{\bm{k}hsa})^* = f^{<e}_{\bm{k}has}$ in the Hermitian conjugate term and defined the real valued purely active rate:
\begin{equation}
    \begin{split}
       \Gamma^{\g e}_{{\rm ZZ,H}\bm{k}h aa} \equiv
         32 G_{\rm F}^2  \, \delta_{e,e'} \delta_{e,-h} 
        \sum_{\{a_i\}} \frac{1}{2|\vec{k}|}  \int &{\rm d PS}_3  
        \; (k^e \cdot p_{2}^{a_2})(p_{1}^{a_1} \cdot p_{3}^{a_3}) 
        \\
    &\times  f^{\g a_1}_{h_1\evec{p}_1aa}
             f^{\l a_2}_{h_2\evec{p}_2aa}
             f^{\g a_3}_{h_3\evec{p}_3aa}.
       \end{split} 
    \label{eq:raterate}
\end{equation}
Here the direct term and interference term give equal contributions and are summed in the rate~\cref{eq:raterate}. As expected, the sterile state has no collision integral in the flavour basis and the off-diagonal terms are damped by rate that is half of the (sum of the sterile and) active rate(s). Further identification of the particle and antiparticle channels proceeds as in the previous example with neutrino-antineutrino scattering.

We have now reduced our initial collision integral with all flavour- and antiparticle mixing effects down to the flavour diagonal limit, where the contact to usual field theoretical methods was easy to make. We also extracted the known structure of the damping terms affecting the coherent flavour evolution in the two-flavour active-sterile mixing case in the UR-limit. We hope that showing these explicit results make it easier for the reader to understand how apply our methods to study any given problem at hand. 

%
\section{General forward scattering potential term}
\label{sec:General_self-energy}
%

We now show how to evaluate the Hermitian self-energy diagrams directly at one-loop level. To be specific, we consider the leftmost bubble diagram in figure~\cref{fig:1loop self-energy}, with an internal neutrino line (we label it by ZB). From this example it should be evident how the other diagrams are computed. We begin by observing that the Hermitian self-energy is equal to the Hermitian part of the $11$-component of the self energy in the CTP indices: $\bar \Sigma_{\rm H} = {\rm He}(\bar \Sigma^{11})$.  We shall evaluate the self-energy and the corresponding forward-scattering coefficient~\cref{eq:WrmH} in the spectral limit, discussed in section~\cref{sec:spectral_limit}. We also first assume that $\bar U_{ij}=\delta_{ij}$. It then is easy to see that the ZB-diagram gives:
\begin{equation}
\bar \Sigma_{{\rm H}il}^{\rm ZB}(k,x) = {\rm He}\left( i\Big(\frac{ig}{2\cos\theta_W}\Big)^2 \int_q 
\gamma^0\gamma^\mu P_{\rm L} iS^{11}_{il}(q,x)\gamma^\nu P_{\rm L} i{\cal D}^{11}_{{\rm Z}\mu\nu}(q-k,x) \right),
\end{equation}
To obtain the most general result, we should derive a non-equilibrium CTP-propagator also for the gauge boson. In most cases of interest however, the gauge bosons appear only as intermediate resonances, which allows to replace ${\cal D}^{11}_{{\rm Z}\mu\nu}$ with the standard vacuum Feynman propagator. For the $S^{11}$-function we use the identity $S^{11} = S^r-S^\l = S_{\rm H} - i{\cal A} - S^\l$. Using the spectral results~\cref{eq:free_spec_fun} and~\cref{eq:cQPA_prop_Wigner_full} we then get
\begin{equation}
iS^{11}_{il}(q,x) 
    =\frac{i\delta_{il}}{\slashed q - m_i + \mathrm{i} \eta}
- \sum_{\th bb'} \big( \delta_{b,-1}\delta_{bb'}\delta_{il} + f_{\vec{q} \th il}^{\l bb'}(t,\evec{x}) \big)  
            \frac{\pi}{\bar\omega_{\evec{q}ij}^{bb'}} 
            D^{bb'}_{\evec{q}\th il} \delta(q_0 - \bar{\omega}_{\vec{q} il}^{bb'}),
\label{eq:S11spectral}
\end{equation}
where we combined part of the spectral function with the Hermitian propagator $S_{\rm H}$ (the principal value propagator in the spectral limit) to extract the vacuum propagator term. It is easy to see that~\cref{eq:S11spectral} reduces to the standard thermal Keldysh propagator in flavour diagonal thermal limit. The vacuum contribution from $iS^{11}_{il}$ is absorbed by the standard renormalization procedure and we only need to consider the second term in~\cref{eq:S11spectral}. Assuming that the energy scales in the problem are small compared to the gauge-boson mass, we can further set $i{\cal D}^{11}_{{\rm Z}\mu\nu} \approx ig_{\mu\nu}/M_{\rm Z}^2$. In this (tadpole) limit $\Sigma_{{\rm H}il}^{\rm ZB}$ is actually independent of $k$, and we can perform the $q_0$ integral trivially to get
\begin{equation}
\bar\Sigma_{{\rm H}il}^{\rm ZB}(x) \approx -\sqrt{2}G_{\rm F} \sum_{\th bb'}     
\int \frac{{\rm d}^3\bm{q}}{(2\pi )^32\bar\omega_{\evec{q}il}^{bb'}} \big( \delta_{b,-1}\delta_{bb'}\delta_{il} + f_{\vec{q} \th il}^{\l bb'}(t,\evec{x}) \big) 
\gamma^0\gamma^\mu P_{\rm L} D^{bb'}_{\evec{q}\th il} \gamma_\mu P_{\rm L} .
\label{eq:S11spectral2}
\end{equation}
To get the forward scattering coefficient one inserts this function to the trace in the formula~\cref{eq:WrmH}. Equation~\cref{eq:S11spectral2} is a very general result however, and to keep the discussion simple, we now specialize to the UR-limit, which is also frequency diagonal. Employing the form~\cref{eq:full_collision_integral_term}, using results~\cref{eq:helicity_simplification,eq:helicity_simplification_2} and computing the resulting simple trace term, one finally gets
\begin{equation}
(\W^{{\rm ZB,H}ee}_{\evec{k}hij})^{l}_{e} \approx \delta_{e,-h} \sqrt{2}G_{\rm F}    
\sum_{b} \frac{1}{2|\bm{k}|}\int \frac{{\rm d}^3\bm{q}}{(2\pi )^32|\bm{q}|} 
\big( \delta_{b,-1}\delta_{il} + f_{\vec{q}-bil}^{\l bb}(t,\evec{x}) \big) 
4e q^b\cdot k^e.
\label{eq:WH_spectral_limit}
\end{equation}
Since the expression in the right hand side of~\cref{eq:WH_spectral_limit} is independent of $j$, we can define, similarly to~\cref{{eq:matter_hamiltonian}}, $\smash{(\W^{{\rm ZB,H}ee}_{\evec{k}hij})^{l}_{e} \equiv (V^{{\rm ZB},e}_{\evec{k}h})_{il}}$. Setting $\bm{q}\rightarrow b\bm{q}$ and using the Feynman-Stueckelberg relation~\cref{eq:FS} along with the equality $\bar f^\g_{\evec{k}hij} = \delta_{ij} - \bar f^\l_{\evec{k}hij}$, we find
\begin{equation}
(V^{\rm{ZB},e}_{\evec{k}h})_{il}(t,\evec{x}) = \delta_{e,-h} \sqrt{2}G_{\rm F}    
\int \frac{{\rm d}^3\bm{q}}{(2\pi )^3} (1-\hat {\bm q}\cdot \hat {\bm k})
\big( f_{\vec{q}-il}^\l(t,\evec{x}) - \bar f_{\vec{q}+il}^\l(t,\evec{x})\big) .
\label{eq:WH_spectral_limit2}
\end{equation}
If the background is isotropic, then the directional term proportional to $\hat {\bm q}\cdot \hat {\bm k}$ vanishes, and we recover the familiar expression~\cite{Enqvist:1990ad,Sigl:1992fn}, written in the vacuum mass eigenbases. In the supernova application, the directional term cannot be neglected however, and the complete structure shown in~\cref{eq:WH_spectral_limit2} should be used. 

It is easy to generalize the above calculation for a general mixing matrix $\bar U_{ij}$. The result is simply that $(V^{{\rm ZB},e}_{\evec{k}h})_{il} \rightarrow (\bar UV^{{\rm ZB},e}_{\evec{k}h}\bar U)_{il}$. This can be further rotated back to the flavour basis using the rotation matrix $U_{i\alpha}$. For example, in the two-flavour active-sterile mixing case discussed above, the flavour space effective potential gets the expected form:
\begin{equation}
(V^{{\rm ZB},e}_{\evec{k}h})_{\alpha\beta} = \left( \begin{array}{cc}
                   (V^{{\rm ZB},e}_{\evec{k}h})_{aa} & 0 \\
                   0 & 0 
                   \end{array}\right),
\label{eq:VZBe}
\end{equation}
where $(V^{{\rm ZB},e}_{\evec{k}h})_{aa}$ is as in equation~\cref{eq:WH_spectral_limit2} with $\smash{f_{\vec{q}-aa}^\l = c^2 f_{\vec{q}-11}^\l + s^2 f_{\vec{q}-22}^\l + cs(f_{\vec{q}-21}^\l + f_{\vec{q}-12}^\l)}$ and similarly for the antiparticle term. Other diagrams can be computed similarly. In particular the tadpole diagram contributes a term  
$\smash{(V^{{\rm ZT},e}_{\evec{k}h})_{il} = \bar U_{il}{\rm Tr}[\bar U(V^{{\rm ZB},e}_{\evec{k}h})]}$ with the same set of approximations. In the above 2-neutrino active-sterile mixing case this leads to the same result as for the ZB-diagram: $(V^{{\rm ZT},e}_{\evec{k}h})_{\alpha\beta} = (V^{{\rm ZB},e}_{\evec{k}h})_{\alpha\beta}$, which is again the expected result.

%
\section{Weight functions}
\label{sec:weight_functions}
%

A central element in our reduction of the non-local Kadanoff--Baym equations to a density matrix equation, was temporal localization, or from the Wigner space point of view, the integration over the frequency variable. The resulting loss of closure required new assumptions about the full correlation function, elaborated in sections~\cref{sec:homogeneous_Ansatz} and~\cref{sec:spectral_limit}. While this approach is clearly successful, one might ask if and to what extend it was unique? It is indeed not, albeit the final equations often do not depend on precise details. The issue is related to how the prior information, or preparation of the system affects its evolution. 

First consider the exact solutions to the KB-equation~\cref{eq:KB_mixed} at formal level. Qualitatively we know that, due to gradient corrections and finite widths as well as flavour and particle-antiparticle mixing, they manifest some intricate structures, which in general are well localized in the phase space. We have solved these structures explicitly in the spectral limit in section~\cref{sec:spectral_limit}.
However, we can never have complete information of any system we observe or describe, and sometimes the resolution of the setup is too poor to discern certain individual structures. That is, we never have access to but some coarse-grained version of the actual system. 

Since by the "system" we basically mean the correlation function, we can formally express the relation between the exact and the observable systems as follows:
\begin{equation}
    S_{ij} (\bar{k}, \bar{x}) \equiv \frac{1}{(2 \pi)^4}\int {\rm d}^4k \, {\rm d}^4x \,
    \mathcal{W}(\bar{k}, \bar{x}; k, x) S_{ij}(k, x),
\label{eq:weighted_average}
\end{equation}
where $\mathcal{W}$ is the weight function encoding the observational resolution. The weight functions can affect any or all variables relevant for the problem. The parameters relevant for this paper were helicity, frequency, 3-momentum, and spatial and temporal coordinates. In the electroweak baryogenesis problem one would be particularly interested in the momentum, the spatial coordinate and the spin perpendicular to the phase transition wall. We only display the weight functions for continuous variables for brevity.

In this language taking the local limit corresponds maximal coarse graining in the frequency variable, or to a statement that there is no prior information about the frequencies. This can be formally described by the weight function 
\begin{equation}
\mathcal{W}(\bar{k}, \bar{x}; k, x) = (2\pi)^3 \delta^3(\Bar{\evec{k}}- \evec{k}) \, \delta^4 (\Bar{x} - x).
\label{eq:simple_weight1}
\end{equation}
This setup is suitable for studying problems including particle-antiparticle mixing, such as particle production, where the particle-and antiparticle mixing shells are widely separated. However, in other problems some different weight functions might be more appropriate. 

The weight function~\cref{eq:simple_weight1} is very simple, consisting of a flat distribution in frequency and delta-functions in 3-momentum and space-time coordinates. These are idealizations of more general functions. The flat weight could, to the same effect, be replaced by a very broad and the delta-functions by very narrow Gaussian distributions. In some cases it can be useful to use such weight functions to enter more detailed prior information on the setup. Indeed, we have seen an example of this already in section~\cref{subsec:particle-antiparticle}, where we integrated out the particle-antiparticle mixing from the general projected master equation~\cref{eq:AH4}. For the parameters we used this procedure is effectively equivalent to using the weight  
%
%
function\footnote{Weierstrass transform on an equation is not exactly equal to using weight function~\cref{eq:simple_weight2} on correlation function in~\cref{eq:weighted_average}. However, our parameters were chosen such that the weight function either left variables intact or averaged them out and in this limit the two are equivalent.}
%
%
\begin{equation}
\mathcal{W}(\bar{k}, \bar{x}; k, x) = (2\pi)^3 \delta^3(\Bar{\evec{k}}- \evec{k}) \, \delta^3 (\Bar{{\bm x}} - \bm{x})\frac{1}{\sqrt{2 \pi}\sigma}\exp(-(t-\bar{t})^2/2\sigma^2).
\label{eq:simple_weight2}
\end{equation}
The same effect could be obtained also by a Wigner space weight function does not have the resolution to erase the flavour structures but erases the information from the particle-antiparticle mixing. For example:
\begin{equation}
   \mathcal{W}(\bar{k}, \bar{x}; k, x) = (2\pi)^3 \delta^3(\Bar{\evec{k}} - \evec{k}) \, 
                        \frac{1}{\sqrt{2 \pi}\sigma_k} \exp(-(k_0-\omega^{a}_{{\bm k}i})^2/2\sigma^2_k)
                        \delta^4 (\Bar{x} - x).
    \label{eq:mixing_weight3}
\end{equation}
Given $2\Delta \omega^{ab}_{{\bm k}ij} \ll \sigma_k \ll \bar\omega^{ab}_{{\bm k}ij}$ this weight function picks the frequency diagonal particle- and antiparticle solutions of all flavours, but suppresses all particle-antiparticle mixing solutions around $k_0 = 2\Delta \omega^{ab}_{{\bm k}ij}$. 

Of course the use of a particular weight function should be motivated by physical arguments, but it is not difficult to see how such motivation would arise in specific setups. Consider for example a laboratory experiment with a neutrino beam. One can usually determine very well what particle or antiparticle flavour states are produced, but their energy and momentum resolution is insufficient to resolve the emitted mass eigenstates. This lack of knowledge imposes the need to integrate exact QKE's over some phase space patch with a weight function, possibly of the form~\cref{eq:mixing_weight3}, with parameters reflecting the experimental resolution. When the experimental resolution is poor compared to spacing of flavour structures, this procedure would lead to our equation~\cref{eq:AH5}, independent of the precise structure of the weight function.

These are just some simple examples of how the prior information imposed on system by integration with a weight function affects the resulting evolution equations. More general weight functions could turn out to be useful tools to study quantitatively how the specific details of neutrino production and detection processes, \eg~the observer-system interference, affects the evolution of the system.  

%

\section{Discussion and conclusions}
\label{sec:discussion}
%

In this article we derived quantum kinetic equations (QKEs) for neutrinos that encompass both the flavour and the particle-antiparticle mixing. Our results include explicit forward scattering terms and collision integrals for the coherent neutrino states. We started from the most general Kadanoff--Baym equations and reduced them, in a set of well defined steps, into a single local density matrix equation~\cref{eq:AH4}, which is valid for arbitrary neutrino masses and kinematics and contains all flavour and particle-antiparticle coherence effects. To our knowledge evolution equations of this generality have not been presented before. We then showed how to consistently integrate out the particle-antiparticle mixing and derived separate (but coupled) equations for particles and antiparticles~\cref{eq:AH5} that are still valid for all kinematic variables. Finally we took the ultra-relativistic (UR) limit of these equations recovering  the familiar form of a density matrix equation~\cref{eq:AH2_UR}.

Our analysis is closely related to earlier work done in refs.~\cite{Her082,Her081,Her083,Her09,Her10,Her11,Fid11,Jukkala:2019slc}, but extending it in many ways. Pivotal elements in our derivation were the careful separation of the pole- and statistical KB-equations and the introduction of the projective representation~\cref{eq:correlator_par}. This unveiled a novel shell structure underlying the mixing phenomenon and allowed expressing the master equation as a set of scalar-valued equations for distribution functions classified according to well defined oscillation frequencies. The only physical assumptions made during the derivation were  the slowly (adiabatically) varying background fields, the validity of the weak coupling expansion and eventually the spectral limit. 

The definition of the collision integrals with all information of the flavour and particle-antiparticle coherences is a very delicate problem, whose importance has been recently emphasized~\cite{Volpe:2023met}. Although some computations in the absence of the particle-antiparticle mixing exist~\cite{Froustey:2020mcq, Bennett:2020zkv,Froustey:2021azz,Xiong:2022vsy, Hansen:2022xza, Capozzi:2018clo} and even some others that do include them~\cite{Her11,Fid11,Jukkala:2019slc}, a complete and comprehensive treatment of the forward scattering terms and collision integrals with the most general mixing structure and arbitrary neutrino masses and kinematics has not existed so far. In ref.~\cite{Blaschke:2016xxt} a formulation of the collision integral is presented in the relativistic limit, but it does not seem convenient for practical purposes. In contradistinction, our derivation includes a simple set of Feynman rules which provide a straightforward and systematic way to compute collision integrals for the flavor- and the particle-antiparticle mixing systems. We showcased the simplicity of our formalism by deriving several simple and/or known limiting cases for the collision integrals. In particular we identified the damping terms in the two-flavour active-sterile mixing case in the UR-limit.

In addition to explicit collision integrals, we also showed how to compute the forward scattering corrections coming from diagrams involving coherent states, i.e. the forward scattering effect arising from a coherent neutrino background. Again we first showed how to obtain the most general structures with both flavour and particle-antiparticle coherences and then took the frequency-diagonal UR-limit of this result, which eventually revealed the familiar structures found using other, less fundamental approaches.

Finally, we briefly discussed how the prior information on the system impacts on its evolution. We pointed out that our localization procedure corresponds to a complete ignorance of the frequency structure of the correlation function and that it is precisely this lack of information that allows for the non-trivial oscillation structure to emerge. However, the perfect localization is an idealization and we sketched how more detailed information of the system could be imposed by the use of specific weight functions. This seems like a promising way to study the observer-system interference in general, beyond the astrophysical and early universe applications.

It has been speculated that neutrino-antineutrino coherence could be relevant in some astrophysical environments or in experiments involving the decay of heavy neutrinos~\cite{Volpe:2023met,Antusch:2020pnn}, but we believe that this is not the case. Because due to their very fast rate the particle-antiparticle oscillations average out completely in the time scales of interest. An exception to the rule could be the leptogenesis problem in the non-resonant regime. In contrast, the particle antiparticle mixing is essential for the particle production problem, \eg~during the (p)reheating phase after inflation, and our most general QKE's~\cref{eq:AH5} are the right tool for studying this problem in the presence of flavour mixing. However, in the astrophysics applications the relevant physics is the CP-violating {\em flavour} mixing in the separate, but possibly strongly coupled (via the forward scattering term) particle and antiparticle sectors. For these systems, that include the core collapse supernovae, nascent neutron stars, compact object mergers and the primordial nucleosynthesis, our frequency diagonal equations~\cref{eq:AH5}, and quite often their UR-limit~\cref{eq:AH2_UR}, are sufficient. 

The same critique applies to some investigations of the CP-violating decays of heavy Majorana neutrinos in collider experiments. Indeed, there has been some interest in the possibility of measuring the heavy neutrino mixing parameters in colliders, because this could give insight into the neutrino mass generation mechanism and eventually into the question if the baryon asymmetry could be explained by the leptogenesis mechanism. To be specific, it has been argued~\cite{Antusch:2020pnn} that the heavy neutrino-antineutrino oscillations could lead to oscillation in the lepton number conserving (LNC) and the lepton number violating (LNV) decay rates, which could be testable at the LHC. Again, the physics behind the phenomenon is {\em not} particle-antiparticle coherence, but the CP-violating flavour mixing of the heavy neutrinos. For these systems our frequency diagonal equations~\cref{eq:AH5} are the ideal tool to use, valid for arbitrary neutrino masses and kinematics.

Our QKE's were derived with only a very few approximations, but there are still some limitations to their applicability. In the local limit, which corresponds to working to the lowest order in the frequency moment expansion in the Wigner space, all truly non-local effects, such as quantum entanglement, are lost. However, we have a rather clear idea about the size of the neglected effects, as they are all encoded in the gradient expansion in the Wigner space KB-equations. If the adiabatic limit in the spatial variables is a reasonable approximation, which often is the case, then all such effects are very small. This type of non-locality could be relevant in some early universe applications however, and it has been studied in the context of the QKE's \eg~in~\cite{Jukkala:2019slc}. Also, when computing the collision integrals and the forward scattering terms we eventually assumed the spectral limit, which neglects the finite width corrections. For the light neutrinos this should be an excellent approximation, but in the case of heavy unstable neutrinos the finite width effects could be interesting. We tried to address this issue by structuring our derivation such that the extension of our analysis to include finite width corrections should be at least in principle evident and it would be interesting to study these issues more in future. Our results should be useful in the study of many interesting problems related to astrophysics, particle physics and physics of the early universe.

%
\section*{Acknowledgements}
%

The work of HP was supported by grants from the Jenny and Antti Wihuri Foundation. We thank Henri Jukkala and Joonas Ilmavirta for many insightful discussions and early collaboration on topics covered in this article. 

%
\appendixtitleon
\appendixtitletocon
\begin{appendices}
%

%
\section{Trace reduction formulae in vacuum eigenbasis}
\label{app:trace_reduction_formulae} 
%

The complete set of Dirac structures consistent with the homogeneity and isotropy of the space
can be chosen as follows.
\begin{equation}
{\rm X} = \{ \mathbbm{1}, \gamma^0, \vec{\gamma} \cdot \hat{\vec{k}}, {\vec{\alpha}} \cdot \hat{\vec{k}}, \gamma^5, \gamma^0\gamma^5, \vec{\gamma}\cdot \hat{\vec{k}}\gamma^5, \vec{\alpha} \cdot \hat{\vec{k}} \gamma^5 \}.
\label{eq:basis1}
\end{equation}
All projections of the effective self-energy functions $\bar \Sigma_{\rm H} = \gamma^0\Sigma_{\rm H}$ appearing in the projected master equation~\cref{eq:AH4} can be reduced to simple combinations of the projected tensors of the form
\begin{equation}
   {\cal O}_{\evec{k}hlji}^{abc} \equiv \Tr[ P_{\vec{k} h il}^{c a} {\cal O} P_{\vec{k} h ji}^{b c}], 
\label{eq:Oproj}
\end{equation}
where ${\cal O}\in {\rm X}$. For all other structures (the labeling follows from the way the operators appear in $\Sigma_{\rm H}$) apart from the pseudoscalar $\gamma^0\gamma^5$ and the contracted tensor $\evec{\gamma}\cdot \hat{\evec{k}} = \gamma^0\frac{1}{2}[\gamma^0,\evec{\gamma}]\cdot \hat{\evec{k}}$, the projection can be written in the form:
\begin{equation}
    {\cal O}_{\vec{k} h lji}^{a b c}  \equiv \frac{1}{2} N_{\vec{k} il}^{ca} N_{\vec{k}ji}^{bc}
     \Big( \frac{A_{\evec{k}hl}^{a}}{(N_{\vec{k} ji}^{bc})^{2}} 
         + \frac{B_{\evec{k}hj}^{b}}{(N_{\vec{k} il}^{ca})^{2}} 
         + C_{\evec{k}hi}^{c} \Big(\frac{1}{(N_{\vec{k} lj}^{ab})^{2}} 
                  - \frac{ab}{\gamma_{\bm{k}l}\gamma_{\bm{k}j}} \Big) \Big),
    \label{eq:sigma_tensor_general}
\end{equation}
where the boost factor $\gamma_\evec{k} = m/\gamma_{\evec{k}}$ and the coefficients $A_{\evec{k}hl}^{a}$, $B_{\evec{k}hj}^{b}$, $C_{\evec{k}hi}^{c}$ and $D_{\evec{k}hl}^{c}$ are listed in table~\cref{tab:reduction_formulae}. 
\begin{table}[h]
\begin{center}
\begin{tabular}{|l|ccc|}
{\rm Operator} & $A_{\evec{k}hl}^{a}$ & $B_{\evec{k}hj}^{b}$ & $C_{\evec{k}hi}^{c}$  \\
\hline
$\mathbbm{1}$; \;\;\phantom{l}$h\evec{\alpha}\cdot\hat{\evec{k}}\gamma^5$  
& $1$ & $1$ & $-1$ \\
$\gamma^0$; \;\; $h\evec{\gamma}\cdot\hat{\evec{k}}\gamma^5$  
& $a/\gamma_{\evec{k}l}$ & $b/\gamma_{\evec{k}j}$ & $c/\gamma_{\evec{k}i}$  \\
$\gamma^5$; \;\; $h\evec{\alpha}\cdot\hat{\evec{k}}$   
& $ahv_{\evec{k}l}$ & $bhv_{\evec{k}j}$ & $-chv_{\evec{k}i}$  \\
\end{tabular}
\end{center}
\caption{Listed are the coefficients in reducted tensor in equation~\cref{eq:sigma_tensor_general} for the spinor structures. }
\label{tab:reduction_formulae}
\end{table}

\noindent Finally, for the Dirac structures $\gamma^0\gamma^5$ and  $h\evec{\gamma}\cdot \hat{\evec{k}}$ the projection tensor is given by
\begin{equation}
    {\cal O}_{\vec{k} h lji}^{a b c}  \equiv \frac{h}{4} N_{\vec{k} il}^{ac} N_{\vec{k}ji}^{bc}
     \Big( - bc \Big( \frac{v_{\evec{k}j}}{\gamma_{\evec{k}i}}
                    + \frac{v_{\evec{k}i}}{\gamma_{\evec{k}j}}  \Big)
        + ca \Big( \frac{v_{\evec{k}i}}{\gamma_{\evec{k}l}} 
                  + \frac{v_{\evec{k}l}}{\gamma_{\evec{k}i}} \Big)
        - ab \Big(  \frac{v_{\evec{k}l}}{\gamma_{\evec{k}j}}
                  - \frac{v_{\evec{k}j}}{\gamma_{\evec{k}l}} \Big) \Big),
    \label{eq:sigma_tensor_alternative}
\end{equation}
where we defined the velocity $v_{\evec{k}i} \equiv |\evec{k}|/\omega_{\evec{k}i}$.

%
\end{appendices}
%

%
\bibliography{references.bib}
%


\end{document}